\documentclass[11pt,a4paper]{article}
\usepackage{cite,mathtools,amsmath,amssymb,color,comment,graphicx}
\usepackage[usenames]{xcolor}
\usepackage[bookmarksopen,colorlinks=true,linkcolor=dark_green,citecolor=dark_red,urlcolor=dark_red,linktocpage=false]{hyperref}
\usepackage[footnotesize]{caption}
\usepackage[height=22.5cm,width=16.5cm,centering]{geometry}

\definecolor{dark_blue}{rgb}{0,0,0.6}
\definecolor{dark_green}{rgb}{0,0.4,0}
\definecolor{dark_red}{rgb}{0.6,0,0}

\newcommand{\bias}{b} 

\def\thefootnote{\fnsymbol{footnote}}

\renewcommand{\thefootnote}{\fnsymbol{footnote}}
\DeclareMathOperator\erfc{erfc}

\begin{document}

\begin{titlepage}

\begin{center}

\hfill DESY 20-029 \\

\vskip 2cm

{\fontsize{15pt}{0pt} \bf
Deformation of the gravitational wave spectrum
}
\vskip 0.5cm
{\fontsize{15pt}{0pt} \bf 
by density perturbations
}

\vskip 1.2cm

{\large
Valerie Domcke, Ryusuke Jinno and Henrique Rubira
}

\vskip 0.4cm

\end{center}
\begin{center}

\textsl{\small Deutsches Elektronen-Synchrotron DESY, 22607 Hamburg, Germany}
\vskip 8pt

\end{center}

\vskip 1.2cm

\begin{abstract}
We study the effect of primordial scalar curvature perturbations 
on the propagation of gravitational waves over cosmic distances. 
We point out that such curvature perturbations deform 
the isotropic spectrum of any stochastic background of gravitational waves of primordial origin through 
the (integrated) Sachs-Wolfe effect.
Computing the changes in the amplitude and frequency of the propagating gravitational wave induced at linear order by scalar curvature perturbations, we show that the resulting deformation of each frequency bin of the gravitational wave spectrum is described by a linearly biased Gaussian with the variance $\sigma^2 \simeq \int d\ln \! k \, \Delta_{\mathcal R}^2$, where $\Delta_{\mathcal R}^2(k)$ denotes the amplitude of the primordial curvature perturbations. The linear bias encodes the correlations between the changes induced in the frequency and amplitude of the gravitational waves.
Taking into account the latest bounds on $\Delta_{\mathcal R}^2$ from primordial black hole and gravitational wave searches, we demonstrate that the resulting ${\mathcal O}(\sigma)$ deformation can be significant for extremely peaked gravitational wave spectra.
We further provide an order of magnitude estimate for broad spectra, for which the net distortion is ${\mathcal O}(\sigma^2)$.

\end{abstract}

\end{titlepage}

\tableofcontents
\thispagestyle{empty}

\renewcommand{\thepage}{\arabic{page}}
\setcounter{page}{1}
\renewcommand{\thefootnote}{$\diamondsuit$\arabic{footnote}}
\setcounter{footnote}{0}

\newpage
\setcounter{page}{1}

\section{Introduction}
\label{sec:Introduction}
\setcounter{equation}{0}

After the first direct  measurement of gravitational waves (GWs) by the LIGO/VIRGO collaboration~\cite{Abbott:2016blz}, a next milestone in GW astronomy will be the discovery of the stochastic gravitational wave background (SGWB). This background will contain unresolved astrophysical sources (such as black hole mergers beyond the resolution limit of the detector) but may also contain cosmological contributions from the very early Universe, such as from inflation, 
preheating,
first-order phase transitions,
and topological defects
(for a review, see \textit{e.g.} Ref.~\cite{Caprini:2018mtu} and references therein). Any GW emitted (or entering the causal horizon) at these early times will be modified by scalar perturbations along its line of sight. In the geometrical optics limit, \textit{i.e.}, in the regime where the wavelength of the GW is much shorter than the wavelength of the scalar perturbation, this will modify the amplitude, frequency and phase of the propagating GW~\cite{Laguna:2009re}. Our goal in this paper is to demonstrate how these effects deform the spectral shape of a SGWB sourced in the Early Universe.

Analogously to the case of CMB photons, mapping an observed GW to the initially emitted GW requires taking into account Doppler, Sachs-Wolfe~\cite{Sachs:1967er}, integrated Sachs-Wolfe (or Rees-Sciama~\cite{Rees:1968zza}) and lensing effects~\cite{Laguna:2009re}. For transient events, the resulting modification of the GW waveform has been studied \textit{e.g.}\ in \cite{Bertacca:2017vod}. For the stochastic background, these effects have been shown to modify and generate anisotropic contributions of the SGWB~\cite{Alba:2015cms,Alba:2017clw,Cusin:2017fwz,Cusin:2017mjm,Bartolo:2019oiq,Bartolo:2019zvb,Bartolo:2019yeu,Bertacca:2019fnt} and to lead to a decoherence of the phase information in the SGWB which becomes relevant when probing the three-point function of GWs~\cite{Bartolo:2018rku,Dimastrogiovanni:2019bfl}.\footnote{Our work is moreover related to but distinct from GW production from large scalar perturbations, which is a (scalar)$^2$ source term in the GW equation of motion~\cite{Acquaviva:2002ud}.} 

Here, given the limited angular resolution of GW detectors, we instead focus on the isotropic component of the GW power spectrum. As we will see, in this case the dominant contribution comes from the Sachs-Wolfe and integrated Sachs-Wolfe effect. The latter modifies only the GW frequency, whereas the former changes both the frequency and amplitude of the propagating GW, leading to a correlated modification of frequency and amplitude. We ignore changes to the phase of the GW since the phase information drops out in the power spectrum. As one of the main results of our work, we provide a master formula encoding the deformation (calculated from the amplitude and frequency changes induced at linear order in the scalar perturbations) of any primordial GW spectrum
by the (integrated) Sachs-Wolfe effect in terms of linearly biased Gaussian kernel. The model-independent linear bias reflects the correlation between amplitude and frequency change inherent to the (integrated) Sachs-Wolfe effect, whereas the variance of the Gaussian is proportional to the integrated scalar curvature power spectrum over the relevant length scales.

The effects in question are of the order (tensor)$\times$(scalar) in the GW equation of motion in the language of scalar-vector-tensor decomposition and are consequently extremely small for a scale invariant primordial scalar power spectrum constrained to the value measured at CMB scales $\Delta_{\mathcal R}^2 \sim 10^{-9}$.\footnote{The resulting effect in the GW power spectrum is of order (scalar)$^2$ for sufficiently broad initial GW spectra, in complete analogy with the CMB. On the contrary, very peaked spectra can experience larger effects, see Sec.~\ref{sec:Results}.} However, little is known about the primordial scalar power spectrum at smaller scales~\cite{Bringmann:2011ut}. A dramatic enhancement may occur \textit{e.g.}\ for specific shapes of the inflaton potential~\cite{Sasaki:2018dmp} or if  particle production during inflation yields an additional source term for scalar perturbations~\cite{Linde:2012bt}. These scenarios have recently received a lot of interest in the context of primordial black hole (PBH) formation, and upper bounds on PBHs can be used to (mildly) constrain the scalar power spectrum at small scales~\cite{Sasaki:2018dmp}. 

In this context, our results can be interpreted in several ways: One the one hand, if the properties of the primordial GW source are well understood by other means, then the comparison with the observed spectrum can be seen as a probe of the primordial scalar power spectrum at the relevant scales. On the other hand, in the more realistic case that the precise source properties are unknown, we compute the degeneracy between the source properties and the propagation effects in the presence of a strongly enhanced scalar spectrum, which could lead to a miss-estimation of the source parameters. Finally, our results can also be used to quantify the smallness of these propagation effects, which become irrelevant if the scalar power spectrum is sufficiently small or if the primordial GW power spectrum is sufficiently broad.\footnote{We caution that in the latter case, our master formula only serves as an estimate for the magnitude of the effect, since we omit ${\mathcal O}(\sigma^2)$ effects in the amplitude and frequency changes. See Sec.~\ref{sec:Results}.} Our results will be particularly relevant for future GW detectors such as LISA~\cite{Audley:2017drz}, the Einstein Telescope~\cite{ET} or DECIGO~\cite{Seto:2001qf}, which will be able to measure the SGWB with high accuracy. 

The organization of this paper is as follows.
In Sec.~\ref{sec:Assumptions} we give an overview of our main results.
We first clarify our setup and then give an intuitive explanation of the results we obtain in the following.
In Sec.~\ref{sec:Basic} we derive our master equations for the deformation of a primordial GW spectrum, with technical aspects relegated to App.~\ref{app:Derivation}.
We in particular demonstrate that the deformation of the GW spectrum is sensitive to the integrated scalar perturbation 
$\int d\ln \! k \, \Delta_{\mathcal R}^2$.
In Sec.~\ref{sec:Constraints} we discuss constraints on this quantity in terms of primordial black hole constraints.
In Sec.~\ref{sec:Results} we illustrate the spectral deformation of a spiky GW spectrum and discuss the limitations of our analysis for broad spectra.
In Sec.~\ref{sec:DC} we discuss implications for future GW experiments and conclude.

\begin{figure}
\centering
\includegraphics[width = 0.7\textwidth]{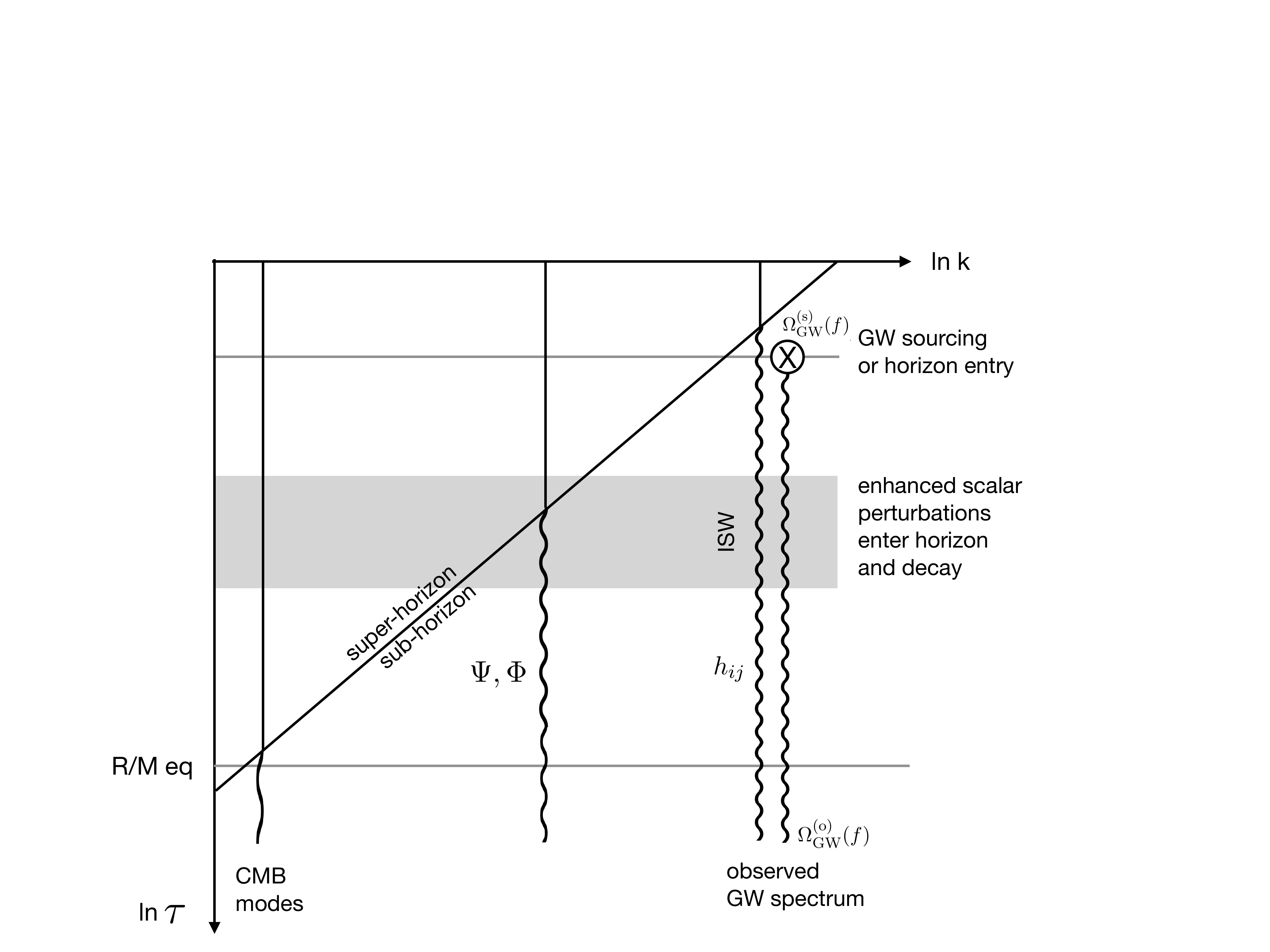} 
\caption{
Sketch of the length and time scales considered in this paper. At large scales (small $k$), the scalar and tensor perturbations are well measured/constrained by the CMB. We study the impact of enhanced intermediate-scale scalar perturbations (labeled $\Phi, \Psi$) on the spectrum of high frequency GWs (labeled $h_{ij}$). Schematically, the gray band indicates the epoch when the decay of the scalar perturbations upon horizon crossing induces a sizable ISW effect on the propagating, sub-horizon GWs. In addition, at the time of sourcing of $h_{ij}$, the scalar perturbations $\Psi,\Phi$ are super horizon, leading to a modification of the locally sourced spectrum through the Sachs-Wolfe effect.
}
\label{fig:sketch2}
\end{figure}

\section{Overview}
\label{sec:Assumptions}
\setcounter{equation}{0}

\subsection{Assumptions and approximations}
\label{subsec:AA}

In this section, we first clarify our assumptions and 
then give an intuitive explanation for the results we obtain in the following sections.
The hierarchy of the length and time-scales relevant to our discussion is sketched in Fig.~\ref{fig:sketch2}.

Our starting point is any high-frequency source of GWs in the early Universe, \textit{i.e.}\ GWs with a wavelength much shorter than the wavelengths probed by the CMB. 
We label the spectrum of these GWs by $\Omega_{\rm GW}^{\rm (s)}(f)$, where ``s" stands for ``source". We assume that these GWs are already inside the horizon when we refer to the spectrum $\Omega_{\rm GW}^{\rm (s)}(f)$ (see also Sec.~\ref{subsec:Explanation}). For a rigorous definition of the sourcing time in the velocity-orthogonal isotropic gauge and the conformal Newtonian gauge, see App.~\ref{app:step1}.

Our goal is to investigate the impact of primordial scalar perturbations (sourced during cosmic inflation) on the propagation of these high-frequency GWs. In particular, we will consider scalar perturbations with a characteristic length scale which is much larger than the wavelength of the GWs but much smaller than the horizon size at CMB decoupling. The former criterion enforces a scale separation which will allow us to treat the propagation of the GWs in the geometrical optics limit. It also ensures that the scalar perturbations enter the horizon only after the GWs have been sourced.
The latter criterion enables us to consider scalar perturbations which are enhanced compared to the scalar two-point correlation function measured in the temperature anisotropies of the CMB. 
Note that this second criterion also ensures that the scalar perturbations enter the horizon in the radiation dominated epoch. We will moreover for simplicity assume that the scalar fluctuations can be described by a Gaussian spectrum. See Fig.~\ref{fig:sketch2} for a visualization of our setup.

As is well known in the context of the propagation of the CMB photons, the (time-dependent) Newtonian potentials $\Phi = \Psi$ encoding the scalar (metric) perturbations induce several effects on the amplitude and frequency of propagating relativistic degrees of freedom, and GWs are no exception. As we will see later, their amplitude and frequency are modified by Doppler, Sachs-Wolfe, integrated Sachs-Wolfe and lensing effects.
In this paper, given the poor angular resolution of GW detectors, we will focus on the isotropic component of the power spectrum of the stochastic GW background.\footnote{
See Refs.~\cite{Alba:2015cms,Alba:2017clw,Cusin:2017fwz,Cusin:2017mjm,Bartolo:2019oiq,Bartolo:2019zvb,Bartolo:2019yeu} for related analysis of the anisotropies of the SGWB.}
In this case, the frequency and amplitude of each individual GW is shifted by ${\mathcal O}(\Phi,\Psi)$, leading to significant net (\textit{i.e.}\ spatially averaged) deformation of strongly peaked GW spectra whereas the deformation is suppressed for sufficiently broad GW spectra.

\begin{figure}
\begin{center}
\includegraphics[width=0.7\columnwidth]{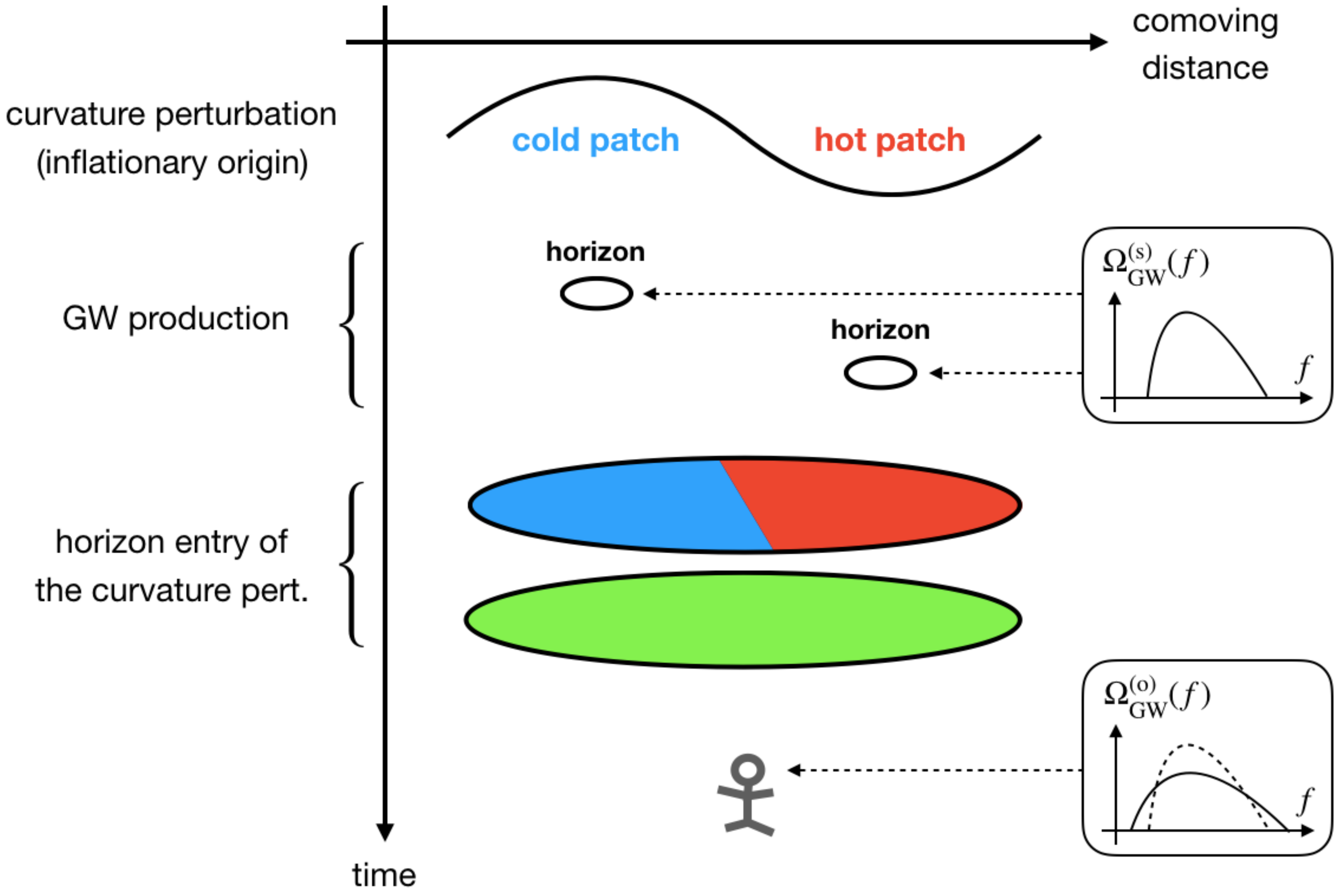}
\caption{\small
A qualitative sketch of the origin of the (integrated) Sachs-Wolfe effect for GWs. The Sachs-Wolfe effect is induced by the super-horizon scalar perturbations at the time of GW production. It can be interpreted as the sum of two effects: the redshift induced when escaping the local Newtonian potential and a local time-delay. The integrated Sachs-Wolfe effect comes into play later, due to the decay of curvature perturbations upon their horizon entry.
}
\label{fig:sketch}
\end{center}
\end{figure}

\subsection{Intuitive explanation of the results}
\label{subsec:Explanation}

Our main result, quantifying the spectral distortion, can intuitively be understood a randomization or smearing of the GW frequency and amplitude induced by the scalar perturbations. This is qualitatively visualized in Fig.~\ref{fig:sketch}: The GW spectrum observed today originates from many different causally disconnected Hubble patches in the early Universe (indicated by the small empty circles).\footnote{
For inflationary GWs, we take ``source time" labeled by ``s'' to be when the relevant GWs are well inside the horizon.
} Assuming adiabatic initial conditions, the scalar perturbations correspond to locally overdense and underdense regions of the Universe (indicated by the blue/red patches). Once these metric perturbations re-enter into the horizon, they quickly decay, leading to a more uniform Universe (indicated by the green patch). In close analogy to the analysis of CMB photons, we can now identify two distinct physical effects:\footnote{
In the language of scalar-vector-tensor decomposition, the effects we consider correspond to the second order terms (tensor)$\times$(scalar) in the equation of motion.
}
\begin{itemize}
\item
\textbf{Sachs-Wolfe effect.} 
At the time when the GWs are sourced, the (at this point in time super-horizon) scalar perturbations distort the sourced GW spectrum. Assuming adiabatic initial conditions, all inhomogeneities can be re-parametrized as a local shift in the time variable, as is characteristic for single-field (single-clock) inflation. In this sense, the event sourcing the GW spectrum (such as the horizon entry of a primordial metric fluctuation or a phase transition) occurs slightly earlier or later in different patches of the Universe, determined by the (super-horizon) scalar perturbations. On top of this, GWs are redshifted according to the value of the Newtonian potential induced by the super-horizon scalar perturbation in a given patch. When computing the isotropic primordial SGWB $\Omega^{\rm (s)}_\text{GW}(f)$ at some fixed time $t^{\rm (s)}$, we have to average over all these patches. The total SW term is the sum of both effects~\cite{Alba:2015cms}.
\item
\textbf{Integrated Sachs-Wolfe effect.} 
When the metric perturbations decay after entering the horizon, the resulting change in the Newtonian potentials induces a frequency shift in the propagating high frequency GWs, described by the integrated Sachs-Wolfe effect. 
\end{itemize}
Adopting the line-of-sight approximation, the comoving frequency of a GW wave is modified by the Sachs-Wolfe and integrated Sachs-Wolfe effect, whereas the amplitude is modified by the Sachs-Wolfe effect as well as by lensing~\cite{Laguna:2009re}. At leading order, these changes can be expressed as linear functions of the Newtonian potentials $\Phi$ and $\Psi$, evaluated at the source time and along the line of sight. Consequently, the dimensionless GW power spectrum $\Delta_h^2$ (see Sec.~\ref{sec:Basic} for the definition) is smeared as the GWs propagate through the inhomogeneous Universe. Gravitational waves arriving from different directions experience a different ``random walk" in frequency and amplitude space, statistically determined by the two-point function of the scalar perturbations on their path. Assuming scalar perturbations of different wavelength to be described by independent Gaussian distributions, we thus expect the net effect to be parametrized as
\begin{align}
\Delta_h^{2 {\rm (o)}}  (f)
&\simeq
\int d\ln \! f'~
\Delta_h^{2 {\rm (s)}} (f')~
K(f,f') \,.
\end{align}
As we will see, the smearing kernel $K$ is given by a linearly biased Gaussian:
\begin{align}
K(f,f')
&=
\frac{1}{\sqrt{2 \pi \sigma^2}}
\left[
1 + \bias \, (\ln \! f - \ln \! f')
\right]
e^{- \frac{(\ln \! f - \ln \! f')^2}{2 \sigma^2}} \,,
\label{eq:Sec2_kernel}
\end{align}
on top of the usual redshift induced by the homogeneous background expansion.
The variance $\sigma^2$ is given by the integrated curvature perturbation 
$\int d\ln \! k \, \Delta_{\mathcal R}^2$ up to an ${\mathcal O}(1)$ factor,
and the linear coefficient $b$ is determined by the correlation between the amplitude and frequency changes,
\begin{align}
\sigma^2
&\simeq
0.91 \times 
\int d\ln \! k~
\Delta_{\mathcal R}^2 \,,
~~~~~~
\bias
\simeq
- 0.52 \,.
\end{align}
Postponing the derivation to Sec.~\ref{sec:Basic} and App.~\ref{app:Derivation},
we can interpret Eq.~(\ref{eq:Sec2_kernel}) intuitively:
The random walk in the frequency gives the Gaussian part of the smearing kernel,
while the linear bias arises because the same scalar perturbations cause a correlated change in the amplitude.

In the following we derive these expressions including the values of the parameters $b$ and $\sigma$, and discuss possible implications for GW observations.

\section{{Deformation of the GW spectrum}}
\label{sec:Basic}
\setcounter{equation}{0}

In this section we summarize the derivation of the spectral deformation.
We direct the readers to App.~\ref{app:Derivation} for the full derivation.

We define GWs by decomposing the metric $ds^2 = g_{\mu \nu} dx^\mu dx^\nu$ as
\begin{align}
g_{\mu \nu}
&= 
\bar{g}_{\mu \nu} + h_{\mu \nu} \,,
\end{align}
with $h_{\mu \nu}$ denoting the two degrees of freedom of the GW tensor~\cite{Isaacson:1967zz,Isaacson:1968zza} (see App.~\ref{app:Derivation}).
Here $\bar{g}_{\mu \nu}$ is the background metric including the scalar perturbations.
We further decompose $h_{\mu \nu}$ (at a fixed $\vec{x}$)
into each frequency $f$, line-of-sight direction $\hat{n}$, and polarization $\lambda$ as 
\begin{align}
h_{\mu \nu} (t, \vec{x})
&=
\int_0^\infty f^2 \, df \int d\Omega~
e^{-2\pi ift}
\sum_{\lambda = +, \times}
h^{(\lambda)} (f, \hat{n}) e^{(\lambda)}_{\mu \nu} (\hat{n})
~+~{\rm c.c.} \,,
\label{eq:h_decomposition}
\end{align}
with the polarization tensor $e^{(\lambda)}_{\mu \nu}$ 
normalized as $e^{(\lambda)}_{\mu \nu} e^{(\lambda') \mu \nu} = \delta_{\lambda \lambda'}$.
The Fourier component $h^{(\lambda)} (f, \hat{n})$ is further decomposed into 
the amplitude $A^{(\lambda)}$ and phase $\phi^{(\lambda)}$:
\begin{align}
h^{(\lambda)} (f, \hat{n})
&= 
A^{(\lambda)} (f, \hat{n}) e^{i \phi^{(\lambda)} (f, \hat{n})} \,.
\end{align}
The power spectrum is defined by the oscillation average over a certain period as
\begin{align}
\langle h^{(\lambda)} (f, \hat{n}) h^{(\lambda')*} (f', \hat{n}') \rangle 
&=
\frac{1}{f^2} \, \delta(f - f') \, \delta^2 (\hat{n}, \hat{n}') \, \delta_{\lambda \lambda'} \, \frac{1}{2} P^{(\lambda)}_h (f) \,,
\label{eq:hh_Ph}
\end{align}
which gives
\begin{align}
\langle h_{\mu \nu} (t, \vec{x}) h^{\mu \nu} (t, \vec{x}) \rangle 
&=
\int d\ln \! f~
4 \pi f^3
\sum_{\lambda = +, \times} \, P_h^{(\lambda)} (f) \,.
\end{align}
We further define the dimensionless power spectrum $\Delta_h^2$ through
\begin{align}
\Delta_h^2 (f)
&=
\sum_{\lambda = +, \times}
\Delta^{2(\lambda)}_h (f)
=
4 \pi f^3
\sum_{\lambda = +, \times}
P_h^{(\lambda)} (f) \,,
\label{eq:Deltah2_Ph}
\end{align}
so that the two-point function of the SGWB is simply given by
the integration of $\Delta_h^2$ over the logarithmic frequency
\begin{align}
\langle h_{\mu \nu} (t, \vec{x}) h^{\mu \nu} (t, \vec{x}) \rangle 
&=
\int d\ln \! f~
\Delta_h^2 (f) \,.
\end{align}
The energy density of GWs per logarithmic frequency interval (normalized by the critical energy density of the Universe) then becomes
\begin{align}
\Omega_{\rm GW}
&=
\int d\ln \! f~
\Omega_{\rm GW} (f)
=
\frac{1}{12 H^2}
\int d\ln \! f~
f^2 \Delta_h^2 (f) \,,
\end{align}
with $H$ denoting the Hubble parameter.
Hereafter we will for notational brevity mostly omit the label for polarization $(\lambda)$.

Our goal is to discuss how the various GW spectra calculated with the FRW background in the literature
are deformed by the existence of long-wave scalar modes.
As explained in the beginning of Sec.~\ref{subsec:Explanation}, 
the GW spectrum we observe is a superposition of GWs coming from different directions,
each of which has experienced different propagation history depending on the realization 
of the scalar mode along the path of propagation.
Since we consider direction-insensitive GW detectors,
this direction-dependent effect results in 
the average of such a scalar-dependent GW spectrum over the scalar mode.\footnote{
Recall that in this paper we consider scalar perturbations with a horizon entry
well before the CMB epoch, which means $l \gtrsim l_{\rm min} \sim 10^2$ for the multipole $l$.
On the other hand, the resolution of future detectors is the order of (sub-)degrees, 
$l_{\rm det} \sim 10 - 100$, at best ({\it e.g.} Ref.~\cite{Cutler:1997ta,Alonso:2020rar}).
Therefore, even such a high-resolution detector sees $\gg (l_{\rm min} / l_{\rm det})^2 \sim 1 - 100$ horizon patches within its resolution.
}
Identifying $\Delta_h^{2 {\rm (s)}}$ with the initial GW spectrum in the comoving gauge
and using logarithmic representation for the frequency, we can show (see App.~\ref{app:Derivation})
\begin{align}
\Delta_h^{2 {\rm (o)}} (\ln \! f)
&=
\left<
e^{2 \,\Delta \! \ln \! A} \,
\Delta_h^{2 {\rm (s)}} 
\left( \ln \! f - \Delta \! \ln \! f \right)
\right>_{\rm ens(s)} \,,
\label{eq:Delta2o_Delta2s_general}
\end{align}
where $\Delta \! \ln \! A$ and $\Delta \! \ln \! f$ are the changes in the logarithmic amplitude and frequency
induced by the scalar modes, and ``ens(s)" denotes an ensemble over the scalar modes.
These quantities denote changes along the line of sight of the propagating GW,
but we can safely replace them with the scalar ensemble average, which thus encodes the sky average over GWs arriving from different directions (see App.~\ref{app:Derivation}).
In this paper we use linear order results for $\Delta \! \ln \! A$ and $\Delta \! \ln \! f$ in terms of the scalar perturbations,
and consider the deformation induced by 
\begin{align}
\Delta_h^{2 {\rm (o)}} (\ln \! f)
&\simeq
\left<
\left( 1 + 2 \Delta \! \ln \! A^{(1)} \right)
\Delta_h^{2 {\rm (s)}} 
\left( \ln \! f - \Delta \! \ln \! f^{(1)} \right)
\right>_{\rm ens(s)} \,,
\end{align}
where the superscript $(1)$ denotes linear order in the scalar perturbations.
As we will see below, this expression accurately describes the deformation of an individual frequency bin or equivalently, the deformation of sufficiently peaked source spectrum.
See discussion in Sec.~\ref{sec:Results} for contributions from higher order terms.

We take the conformal Newtonian gauge for the scalar modes and expand the background metric  
(including the scalar perturbations) as
\begin{align}\label{eq:metric}
\bar{g}_{\mu \nu} dx^\mu dx^\nu
&= 
- a^2 (1 + 2\Phi) d\tau^2 + a^2 (1 - 2\Psi) \delta_{ij} dx^i dx^j \,,
\end{align}
and consider the effect of $\Phi$ and $\Psi$ on the propagation of GWs with infinitesimal amplitude.
Such propagation effects of GWs in inhomogeneous background are discussed in \textit{e.g.} 
Refs.~\cite{Laguna:2009re,Contaldi:2016koz,Cusin:2017fwz,Cusin:2017mjm,Bertacca:2017vod}.
Ref.~\cite{Laguna:2009re} discusses the amplitude and frequency change for point-source GWs, 
and we partly use their results. 
However, in our setup we need special care concerning the initial time slice as we explain below.
For this purpose we define two eras (see also Fig.~\ref{fig:gauge} in App.~\ref{app:Derivation}):
\begin{itemize}
\item
Sourcing time (labeled by ``s"):
\\
This refers to the era when the GWs of our interest become sufficiently sub-horizon.
For GW production mechanisms after inflation, this is the time when the sub-horizon GWs are actually produced,
while for inflationary GWs this is the time when all the relevant modes enter the horizon and become sufficiently sub-horizon.

\item
Observation time  (labeled by ``o"):
\\
This refers to the present time when we observe the GWs.
As we explain later, we neglect the proper velocity of the observer and scalar perturbations around this time.
\end{itemize}
We use the result of Ref.~\cite{Laguna:2009re} to estimate the effect due to the GW propagation from a given
sourcing time {\it in the conformal Newtonian gauge} to the observation time, 
while we separately need to take account of the effect of local time shifts modifying the sourcing time. 
We will label the former with ``prop'' while we will refer to second one as ``init''.
The situation is exactly the same as the Sachs-Wolfe effect in the CMB:
the well-known result for the temperature change 
$(\Delta T / T)_{\rm SW} = (1/3) \Phi_{\rm LS} = \Phi_{\rm LS} - (2/3) \Phi_{\rm LS}$
(in the matter-dominated Universe) can be interpreted as the sum of 
the local Newtonian potential on the last scattering (LS) surface which the photon has to climb out of, $(\Delta T / T)_\text{prop} = \Phi_{\rm LS}$, 
and the local time shift with respect to the time coordinate of conformal Newtonian gauge, $(\Delta T / T)_\text{init} = - (2/3) \Phi_{\rm LS}$~\cite{White:1997vi}.

Similarly, the effect on the amplitude and frequency of the GWs can be decomposed as 
\begin{align} 
&\Delta \! \ln \! A^{(1)}
\equiv 
\left[
\ln (A_{\rm s} / A_{\rm o}) - 1
\right]^{(1)}
= (\Delta \! \ln \! A^{(1)})_{\rm init} + (\Delta \! \ln \! A^{(1)})_{\rm prop} \,,
\\
&\Delta \! \ln \! f^{(1)}
\equiv 
\left[
\ln (f_{\rm s} / f_{\rm o}) - 1
\right]^{(1)}
= (\Delta \! \ln \! f^{(1)})_{\rm init} + (\Delta \! \ln \! f^{(1)})_{\rm prop} \,.
\end{align}
Note again that the superscript (1) indicates first order in the scalar perturbations.
Also note that we are working in conformal coordinates, factoring out the trivial redshift in the FRW background.

Now we discuss the two effects in turn (see App.~\ref{app:Derivation} for details). The effect of the local time shift at the sourcing time is given by
\begin{align}
(\Delta \! \ln \! A^{(1)})_{\rm init}
&=
-\frac{1}{2} \Phi_{\rm s},
\label{eq:DeltalnA1}
\\
(\Delta \! \ln \! f^{(1)})_{\rm init}
&=
-\frac{1}{2} \Phi_{\rm s} \,.
\label{eq:Deltalnf1}
\end{align}
This is the same as the $-(2/3) \Phi_{\rm s}$ term in CMB as explained above,
except that we consider radiation domination $a \propto t^{1/2}$ instead of matter domination $a \propto t^{2/3}$.

The effect due to propagation can be calculated in the geometric optics limit. 
The relative changes in the comoving amplitude and comoving frequency to linear order are given by~\cite{Laguna:2009re}
\begin{align}
(\Delta \! \ln \! A^{(1)})_{\rm prop}
&=
(\Psi_{\rm o} - \Psi_{\rm s}) \,,
\label{eq:DeltalnA2}
\\
(\Delta \! \ln \! f^{(1)})_{\rm prop}
&=
\hat{n} \cdot (\vec{v}_{\rm o} - \vec{v}_{\rm s}) 
- (\Phi_{\rm o} - \Phi_{\rm s})
+ \int_{\lambda_{\rm s}}^{\lambda_{\rm o}} d\lambda~
\partial_\tau (\Phi + \Psi) \,,
\label{eq:Deltalnf2}
\end{align}
Here $\lambda$ is the affine parameter along the path of GWs,
which takes $d/d \lambda = \partial_\tau - n^i \partial_i$ at the unperturbed level.
Note that this unperturbed relation is enough for our purpose, since the integrand in Eq.~\eqref{eq:Deltalnf2} is 
already first order in the perturbations.

In Eq.~(\ref{eq:DeltalnA2}) the term on the right-hand side encodes the Sachs-Wolfe effect,
while we neglected the lensing term.
This is because the lensing is just a rearranging effect of the propagation direction,
and does not affect the isotropic part of the stochastic GW background (see App.~\ref{app:Derivation}).
Moreover, since we are interested in the distortion of the GW spectrum we can ignore $\Psi_{\rm o}$, 
the Newtonian potential at the position of the observer, since this is universal for all GWs, independent of their frequencies.

In Eq.~(\ref{eq:Deltalnf2}) the three terms correspond to 
Doppler, Sachs-Wolfe, and integrated Sachs-Wolfe effects, respectively.
The vector $\vec{v}$ is the spatial component of the fluid velocity $u^\mu = (1 - \Phi, \vec{v})/a$. 
Among the terms in the right hand side of Eq.~(\ref{eq:Deltalnf2}), 
we will neglect  $\Phi_{\rm o}$ and $\vec{v}_{\rm s}$. The former is again universal for all GWs and, in
the cosmic rest frame, the latter is suppressed by the factor of (wavelength)$/$(horizon size), which we assumed to be $\gg 1$ at the sourcing time. Moreover, we will in the following neglect the effect of the observer's velocity $\vec{v}_{\rm o}$. In the case of the CMB, this term (describing the motion of the observer with respect to the cosmic rest frame) gives rise to the observed CMB dipole with $|\vec{v}_{\rm o}| \sim 10^{-3}$. Here we assume that when relevant, the angular resolution of future GW detectors will be sufficient to isolate this dipole contribution~\cite{Allen:1996gp,Kudoh:2005as}.

Now, the relevant terms are
\begin{align}
\Delta \! \ln \! A^{(1)}
&\simeq
- \Psi_{\rm s} - \frac{1}{2} \Phi_{\rm s} \,,
\label{eq:DeltalnA_approx}
\\
\Delta \! \ln \! f^{(1)}
&\simeq
\frac{1}{2} \Phi_{\rm s}
+ 
\int_{\lambda_{\rm s}}^{\lambda_{\rm o}} d\lambda~
\partial_\tau (\Phi + \Psi) \,.
\label{eq:Deltalnf_approx}
\end{align}
We impose $\Phi = \Psi$, assuming that the anisotropic stress is negligible. We also recall that the small-scale scalar perturbations in question enter the horizon during the radiation-dominated epoch.
Then the time evolution of the scalar field is given by~\cite{BaumannCosmology}
\begin{align}
\Phi (\tau, \vec{k})
&=
\Psi (\tau, \vec{k})
= 
\frac{2}{3} {\mathcal R}_{\rm pr} (\vec{k}) T (k\tau) \,,
~~~~
T (k\tau)
=
\frac{9}{k^2 \tau^2}
\left[
\frac{\sin (k \tau / \sqrt{3})}{k \tau / \sqrt{3}} - \cos(k \tau / \sqrt{3})
\right] \,,
\label{eq:RT}
\end{align}
and ${\mathcal R}_{\rm pr} (\vec{k}) = {\mathcal R} (\tau_{\rm pr}, \vec{k})$ is the curvature perturbation outside the horizon. 
We assume that the scalar perturbations $\Phi$ and $\Psi$ originate from the Gaussian primordial 
curvature perturbation ${\mathcal R}_{\rm pr}$.
Its power spectrum is defined as
\begin{align}
\left< \mathcal{R} (\vec{k}) \mathcal{R}^* (\vec{k}') \right>_{\rm ens(s)}
&= 
(2\pi)^3 \delta^3 (\vec{k} - \vec{k}') 
\frac{2\pi^2}{k^3} \Delta_{\mathcal{R}}^2 (k) \,.
\label{eq:p_def}
\end{align}
Now, since $\Delta \! \ln \! A$ and $\Delta \! \ln \! f$ have a nonvanishing correlation, we rotate the basis to eliminate it:
\begin{align}
\left(
\begin{matrix}
\Delta \! \ln \! A^{(1)}
\\[1ex]
\Delta \! \ln \! f^{(1)}
\end{matrix}
\right)
&=
\left(
\begin{matrix}
c & - s
\\[1ex]
s & c
\end{matrix}
\right)
\left(
\begin{matrix}
\Delta_1 \\[1ex]
\Delta_2
\end{matrix}
\right) \,,
\end{align}
with $c \equiv \cos \theta$ and $s \equiv \sin \theta$, 
and $\theta$ is chosen so that $\Delta_1$ and $\Delta_2$ satisfy
\begin{align}
\left<
\Delta_1 \Delta_2
\right>_{\rm ens(s)}
&=
0 \,.
\end{align}
This condition gives the value of $\theta$ in terms of the variances of $\Delta \! \ln \! A^{(1)}$ and $\Delta \! \ln \! f^{(1)}$.
Numerically we find $\theta \simeq -0.696$, see App.~\ref{app:Derivation}. 
Note that this number does not depend on any details of the scalar or tensor power spectrum, but is an intrinsic property the GW propagation encoded in Eqs.~\eqref{eq:DeltalnA_approx} and \eqref{eq:Deltalnf_approx}.
After performing the Gaussian integrations with respect to $\Delta_1$ and $\Delta_2$,
we find that the GW spectrum is deformed by the kernel (see App.~\ref{app:Derivation}),
\begin{align}
K (f, f')
&\simeq
\frac{1}{\sqrt{2 \pi \sigma^2}}
\left[
1 + \bias \, (\ln \! f - \ln \! f')
\right]
e^{- \frac{(\ln \! f - \ln \! f')^2}{2 \sigma^2}} \,,
\label{eq:kernel}
\end{align}
as
\begin{align}
\Delta_h^{2 {\rm (o)}} (f)
&=
\int d\ln \! f'~
\Delta_h^{2 {\rm (s)}} (f')~
K (f, f') \,.
\label{eq:Delta2o_Delta2s}
\end{align}
In terms of the GW energy fraction, $\Omega_{\rm GW} \propto f^2 \Delta_h^2$, we have
\begin{align}
\Omega_{\rm GW}^{\rm (o)} (f)
&=
\int d\ln \! f'~
\Omega_{\rm GW}^{\rm (s)} (f')~
(f / f')^2 K (f, f') \,.
\label{eq:Omegao_Omegas}
\end{align}
Here the variance $\sigma$ and linear coefficient $\bias$ are given by
\begin{align}
\sigma^2
&=
s^2 \sigma_1^2 + c^2 \sigma_2^2 \,,
~~~~~~
\bias
\equiv
\frac{2c s (\sigma_1^2 - \sigma_2^2)}{s^2 \sigma_1^2 + c^2 \sigma_2^2} \,,
~~~~~~
\sigma_1^2
=
\left< \Delta_1^2 \right> \,,
~~~~~~
\sigma_2^2
\equiv
\left< \Delta_2^2 \right> \,.
\end{align}
Numerically we find
\begin{align}
\sigma^2
&\simeq
0.91 \times 
\int d\ln \! k~
\Delta_{\mathcal R}^2 \,,
~~~~~~
\bias
\simeq
- 0.52 \,.
\label{eq:smearing_numerical}
\end{align}
We study implications of Eqs.~(\ref{eq:Delta2o_Delta2s}) and (\ref{eq:Omegao_Omegas}) 
on the deformation of the GW spectrum.
We emphasize that to linear order in the scalar perturbations the results of this section, including the numerical prefactors in Eq.~\eqref{eq:smearing_numerical} 
are generic consequences of the SW and ISW effects and in particular apply (frequency bin by frequency bin) to any cosmological GW spectrum.

Note that since $\langle \Psi \rangle = 0 = \langle \Phi \rangle$, we expect the resulting deformation of $\Delta_h^2$ and $\Omega_\text{GW}$ to be ${\mathcal O}(\sigma^2)$. An exception to this occurs for initial GW spectra with a width smaller than $\sigma$, since in this case the width of the observed spectrum is to leading order set by the induced frequency change $\Delta \!\ln \!f$, and is consequently ${\mathcal O}(\sigma)$. See Sec.~\ref{sec:Results} for details.

\section{Constraints on the curvature perturbations}
\label{sec:Constraints}

As we saw in the previous section, the variance for the spectral broadening (\ref{eq:smearing_numerical}) 
is given by the integrated scalar power spectrum.
The latter is measured by the CMB at large scales and is constrained by  primordial black hole (PBH) bounds on smaller scales.
In this section we review the allowed parameter space for the scalar perturbations.

The present-day mass function of the PBH -- the fraction of dark matter contained in PBH
per logarithmic bin of mass
-- can be parametrized as \cite{Sasaki:2018dmp}
\begin{align}
f_{\rm PBH}(M) \equiv \frac{1}{\rho_\text{DM}}\frac{d\rho_{\rm PBH}}{d\ln{M}} \approx \left(\frac{\beta}{6.6\cdot10^{-9}}\right) \left( \frac{\gamma}{0.2} \right)^{\frac{1}{2}}\left( \frac{106.75}{g_{\ast}}\right)^{\frac{1}{4}}\left(\frac{M_\odot}{M}\right)^{\frac{1}{2}} \,,
\label{eq:fPBH}
\end{align}
in which $g_{\ast}$ is the effective number of light degrees of freedom, $\rho_\text{DM}$ is the energy density of dark matter today and $\gamma$ is the fraction of the horizon mass (at the time of PBH formation) which collapsed into a BH,
\begin{align}
M = \gamma M_H  = \gamma \left( \frac{3 H^2}{8 \pi G} \right) \left( \frac{4 \pi}{3} R_H^3 a^3\right) = \frac{\gamma}{2G}aR_H \,.
\end{align}
The last equality clearly displays the one-to-one relation between the PBH mass the comoving horizon radius $R_H = 1/(aH)$. $\gamma$ is analytically calculated to be $0.2$\cite{Carr:1975qj} and $G$ denotes Newton's constant. 
Assuming a gaussian overdensity field $\delta = (\rho-\bar{\rho})/\bar{\rho}$,
the fraction of the energy density collapsed into PBHs at the time of formation, $\beta$, is given by \footnote{$f_\text{PBH}$, the fraction of PBH today, is simply a rescaling of $\beta$. The $M^{-1/2}$ dependence in Eq.~\eqref{eq:fPBH} takes into account the fact that PBH density scales as matter, whereas the total energy density is diluted as radiation. }
\begin{align} \label{eq:beta}
\beta(M) = \int_{\delta_c}^\infty \frac{d\delta}{\sqrt{2\pi}\sigma_\delta}e^{\delta^2/2\sigma_\delta^2} = \frac{1}{2}\erfc\left(\frac{\delta_c}{\sqrt{2}\sigma_\delta}\right)  \,,
\end{align}
where $\delta_c$ is the critical density above which local overdensities collapse into black holes, measured in simulations as $0.5$ \cite{Musco:2018rwt} and $\sigma_\delta$ is the variance of the overdensity field smoothed by the gaussian filter $ W(k,R_H) = \exp{\left[-(k R_H)^2/2\right] } $ defined as\footnote{There are several points regarding the need for a filter. First, $\delta (x)$ is implemented as a random field,
which is neither continuous nor differentiable \cite{Ando:2018qdb,Kalaja:2019uju}. Second, once we are dealing with scale dependent initial curvature perturbations, the variance of the overdensity will depend on the perturbation mode
entering the horizon. In that sense, the filter ensures the correct probability density function for the field on the horizon scales. For a detailed discussion about the need of a filter, see \cite{Kalaja:2019uju}. Ref.~\cite{Kalaja:2019uju} also highlights that an additional filtering at scales outside the horizon is needed, which is beyond the scope of the present work.}
\begin{align}\label{eq:sigma_def}
\sigma_\delta^2(R_H) = \int_0^\infty d\ln{k}\, W^2(k,R_H)\Delta^2_{\delta}(k) \,,
\end{align}
with $\Delta^2_{\delta}(k)$ being the power spectrum of the overdensity, defined analogously to equation~\eqref{eq:p_def}. One can relate the matter fluctuations with the curvature perturbations through the Poisson equation
\begin{align}
\Delta^2_{\delta}(k) = \left(\frac{4k^2}{9a^2H^2}\right)^2\Delta^2_{\mathcal R}(k) \,.
\end{align}
Note that, taking all of this together, the PBH fraction in Eq.~\eqref{eq:fPBH} or \eqref{eq:beta} depends exponentially on the amplitude of fluctuations $\Delta^2_{\mathcal R}$.

Now we proceed to calculating the bounds for two types of initial conditions for $\Delta^2_{\mathcal R}$: one monochromatic, defined as a Dirac delta at a scale $k_{\ast}$ and one scale invariant spectrum in an interval $\left[ k_{\rm min},k_{\rm max}\right]$. To translate the constraints from $f_{\rm PBH}(M)$ to the amplitude of the primordial spectrum ($A_{s}$ and $k_{\ast}$ for the monochromatic and $A_\theta$ and $k_{\rm max}/k_{\rm min} $ for the scale invariant case, as defined below), we follow \cite{Inomata:2017okj} and use the criteria
\begin{align} \label{eq:translate_bounds}
\int d\ln \! M~
\frac{f_{\rm PBH}(M)}{f_{\rm obs}(M)} \leq 1 \,,
\end{align}
with $f_{\rm obs}$ being the observational constraints, which we extracted from \cite{Inomata:2017okj}.

\begin{figure}[t]
\centering
  \includegraphics[width=0.45\textwidth]{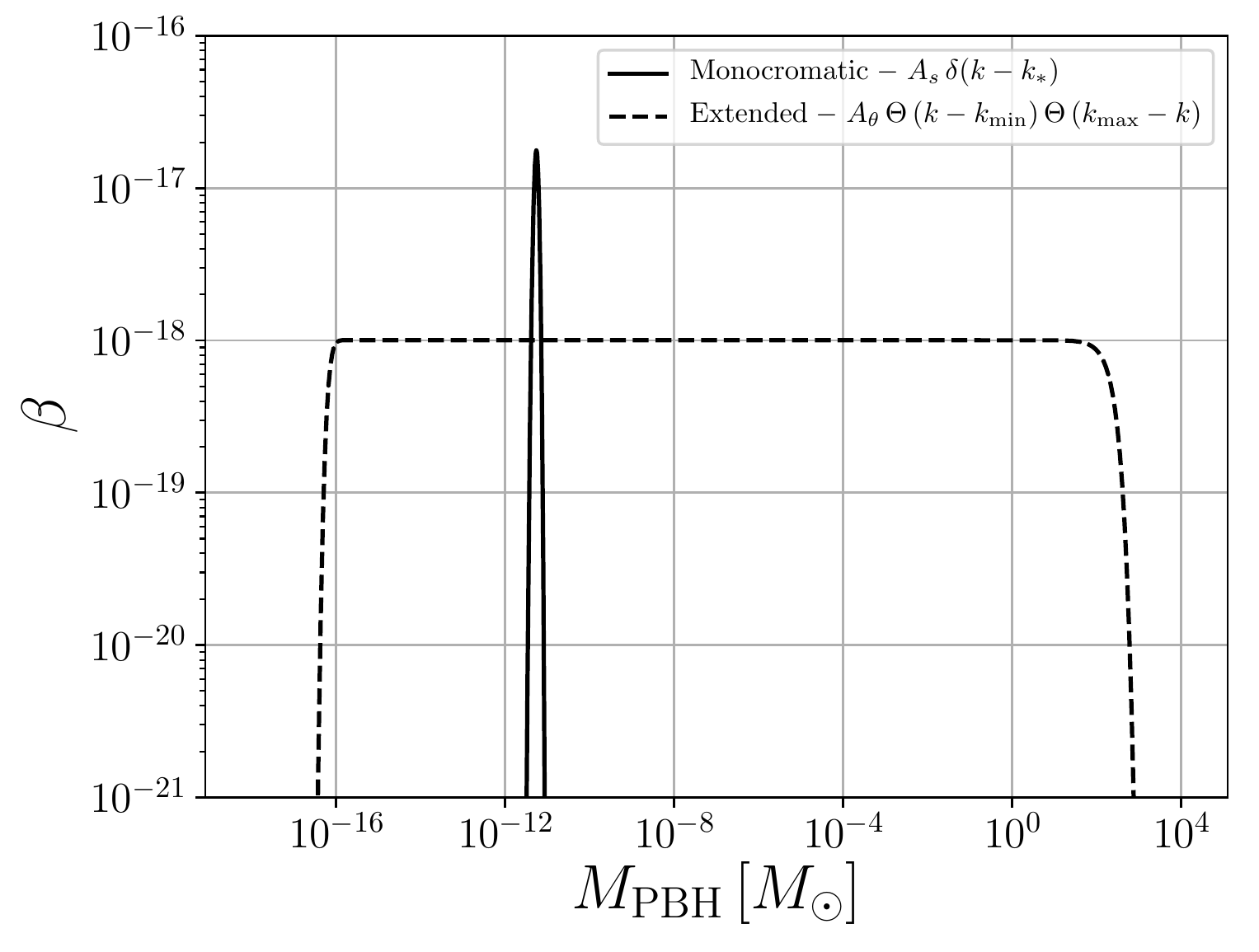}  
  \includegraphics[width=0.45\textwidth]{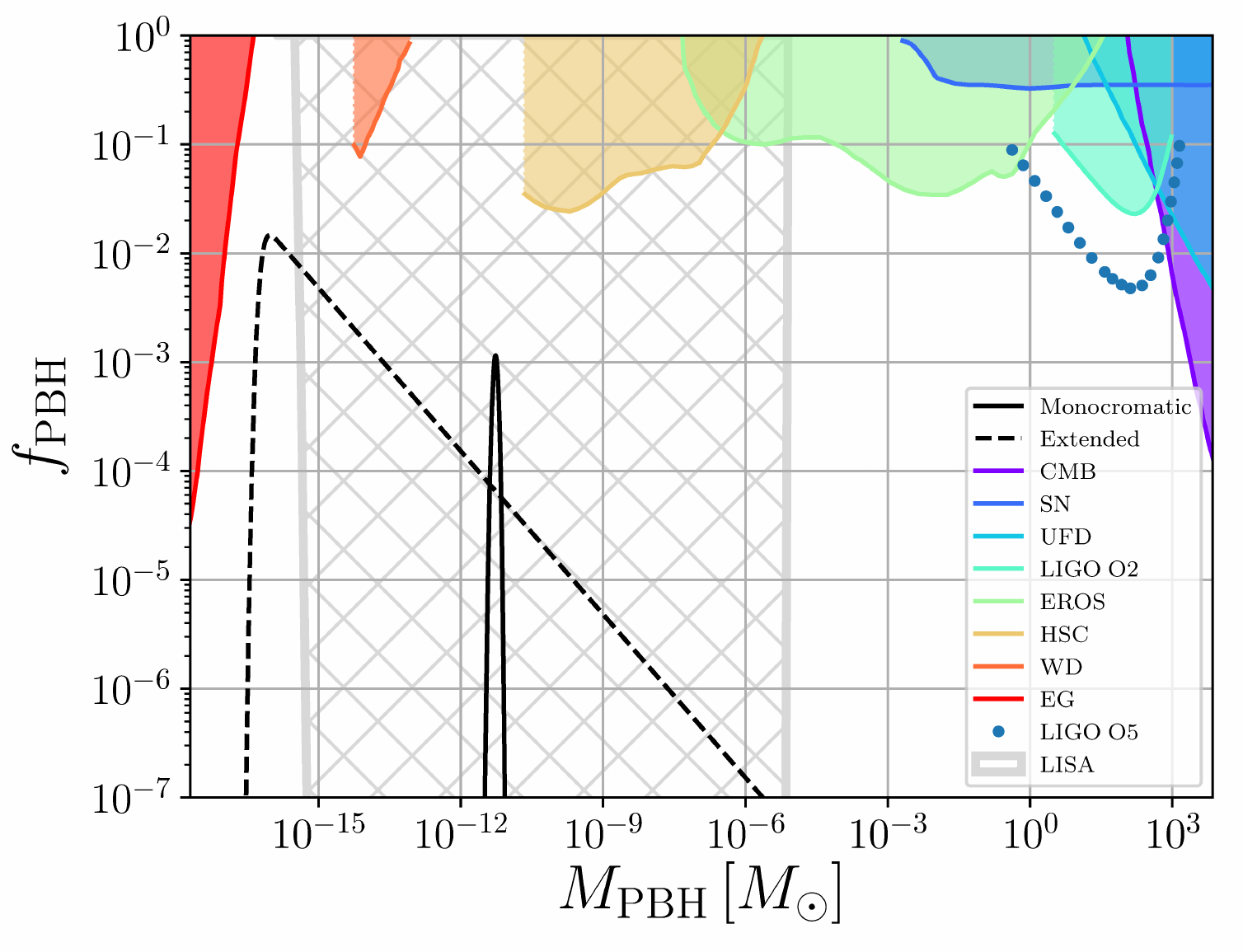}  
\caption{\label{fig:beta}
Exemplary spectra of primordial black holes. 
\textit{Left panel:} 
fraction of energy $\beta$ collapsed into PBHs at the time of formation. For the monochromatic initial condition (with $ A_s = 0.033$, $k_{\ast} = 2\times 10^{12}$ Mpc$^{-1}$) $\beta$ is strongly peaked; for a Heaviside initial condition ($A_\theta = 0.033$, $k_{\rm min} = 10^{5}$ Mpc$^{-1}$ and $k_{\rm max}/k_{\rm min} = 10^{10}$) it is scale invariant in between the threshold masses. 
\textit{Right panel:}
present fraction of PBH $f_{\rm PBH}$. For the PBH constraints, we use \cite{Inomata:2017okj} (with HSC constraints cut following \cite{Katz:2018zrn}). For the supernovae bounds (SN) we used \cite{Zumalacarregui:2017qqd} and we also included LIGO O2 constraints from \cite{Raidal:2018bbj}. The dotted curve displays the expected LIGO O5 \cite{Ali-Haimoud:2017rtz} bound, while the gray hatched band is the expected bound by LISA \cite{Byrnes:2018txb}.}
\end{figure}

\subsubsection*{Monochromatic Gaussian scalar fluctuations}

Monochromatic PBHs are generated by a Dirac delta function 
\begin{align}
\Delta^2_{\mathcal R}(k) = A_s k_{\ast}\delta\left(k-k_{\ast} \right) \,,
\end{align}
for which it is possible to calculate the variance of the overdensity field at a scale $R_H$ analytically through Eq.~\eqref{eq:sigma_def} as\footnote{Notice that even though the initial condition is a Dirac delta, the filter effect leads to a gaussian variance. The choice of the filter of course introduces an additional uncertainty in the prediction. Different filter effects in the PBH constraints are discussed in \cite{Ando:2018qdb}.}
\begin{align}
\sigma_\delta^2(R_H) = \frac{16}{81} A_s k_{\ast}^4R_H^4 \exp{\left[-\left(k_{\ast} R_H\right)^2 \right] } \,.
\label{eq:sigma_delta_mono}
\end{align}
For this case, the variance of the kernel smearing the GW spectrum, see Eq.~\eqref{eq:smearing_numerical},
is
\begin{align} \label{eq:sigma_mono}
\sigma^2 \simeq 0.91 \, A_s \,.
\end{align}
Notice that while $\sigma^{2}_{\delta}$ denotes the variance of the density field, $\sigma^{2}$ with no subscript refers to the broadening of the GW spectrum due to line-of-sight distortions.

\subsubsection*{Extended mass PBH mass function / scale invariant scalar fluctuations}

As a second exemplary case, we consider a scale invariant curvature perturbation spectrum within an interval $\left[ k_{\rm min},k_{\rm max}\right]$
\begin{align}
\Delta^2_{\mathcal R}(k) = A_\theta \, \Theta\left(k-k_{\rm min} \right)\Theta\left(k_{\rm max}-k \right) \,.
\end{align}
This yields for the variance of the overdensity field at a scale $R_H$
\begin{align}
\sigma_\delta^2(R_H) = \frac{16}{81}  \frac{A_\theta}{2}\left( e^{-k_{\rm min}^2R_H^2}(1+k_{\rm min}^2R_H^2) - e^{-k_{\rm max}^2R_H^2}(1+k_{\rm max}^2R_H^2) \right) \,.
\end{align}
For this case, the variance of the kernel smearing the GW spectrum is given by
\begin{align}\label{eq:sigma_ext}
\sigma^2 = 0.91 A_\theta \ln{\left(k_{\rm max}/k_{\rm min}\right)} \,.
\end{align}

In the left panel of Fig.~\ref{fig:beta}, we display the fraction $\beta$ of energy density that collapsed into PBH calculated through Eq.~\eqref{eq:beta} for both the Heaviside and delta function initial conditions. In the right panel, we display the PBH dark matter fraction today as predicted by both models, together with different PBH bounds. For both plots we chose the parameter examples $ A_s = 0.033$, $k_{\ast} = 2\times 10^{12}$ Mpc$^{-1}$ for the monochromatic case and $A_\theta = 0.033$, $k_{\rm min} = 10^{5}$ Mpc$^{-1}$  and  $k_{\rm max}/k_{\rm min} = 10^{10}$ for the Heaviside case.

\begin{figure}[t]
\centering
  \includegraphics[width=0.45\textwidth]{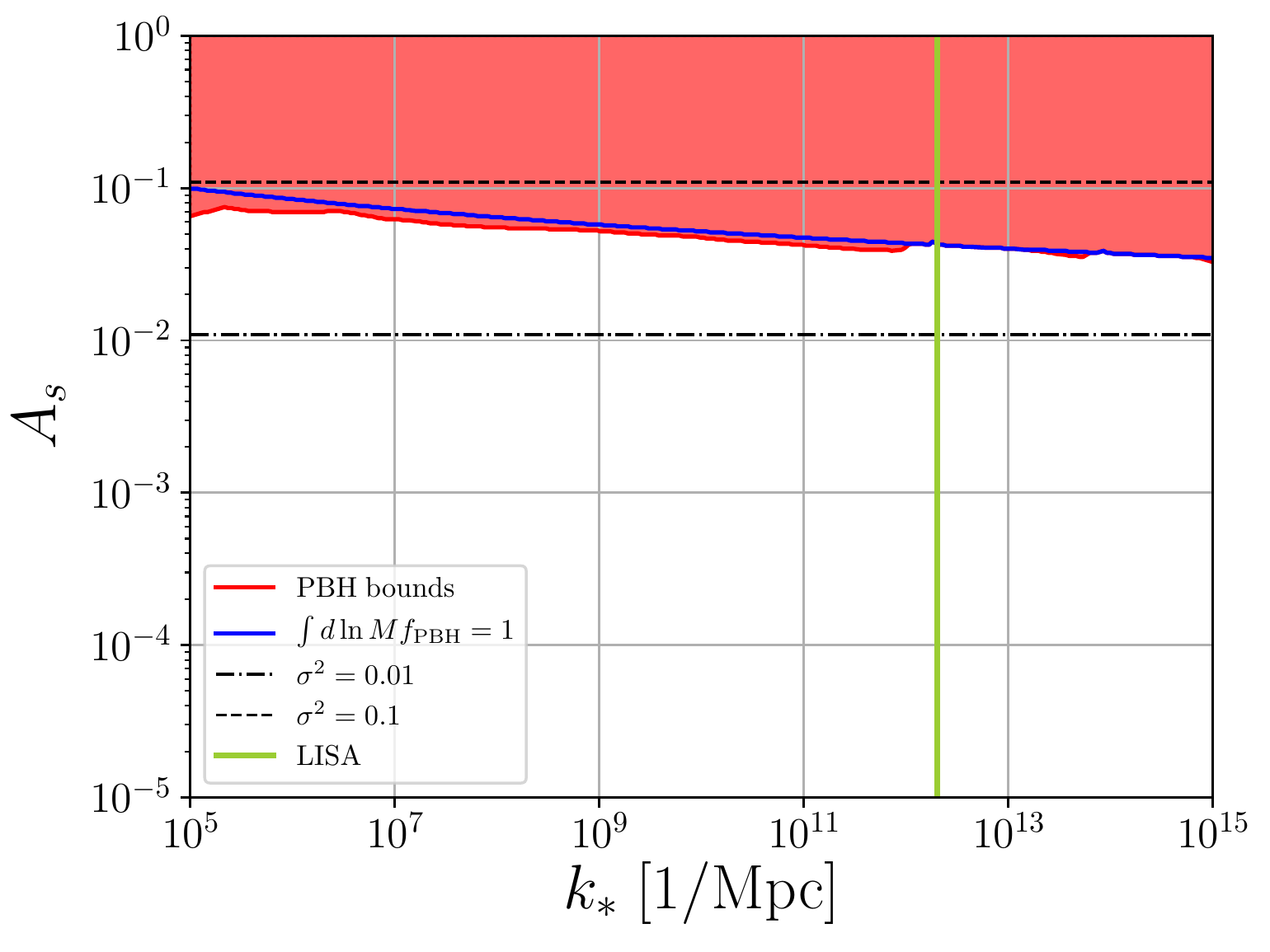}  
  \includegraphics[width=0.45\textwidth]{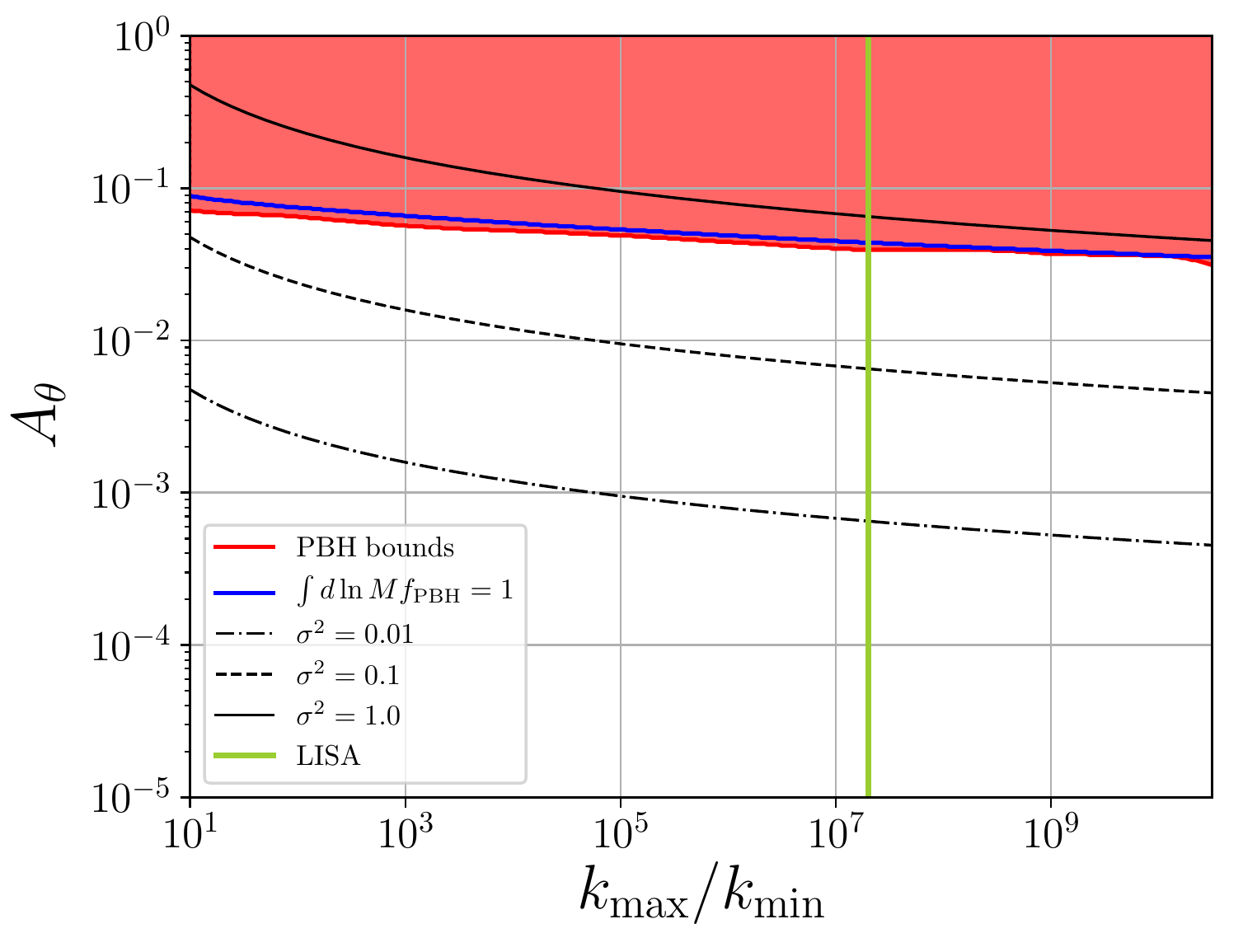}  
\caption{\label{fig:ic_constraint}
Constraints on the amplitude of the primordial curvature power spectrum $\Delta_{\mathcal R}^2$ and on the resulting smearing of the GW power spectrum induced by the (integrated) Sachs-Wolfe effect, encoded in the width of the smearing kernel $\sigma^2$. Left panel: monochromatic primordial curvature power spectrum peaked at $k_*$. Right panel: Extended primordial curvature power spectrum between $k_\text{min}$ and $k_\text{max}$ with $k_{\rm min}$ fixed to $10^{5}$ Mpc$^{-1}$.}
\end{figure}

In order to translate the constraints on $f_{\rm PBH}$ to the parameter space of our two exemplary models, we use equation~\eqref{eq:translate_bounds}. In the left panel of Figure~\ref{fig:ic_constraint}, we explore the parameter space $\{ A_s, k_{\ast} \}$ for the delta function initial condition. In the right panel, we explore the parameter space $\{ A_s, k_{\rm max}/k_{\rm min} \}$ for the Heaviside spectrum fixing $k_{\rm min} = 10^{5}$ Mpc$^{-1}$ (this corresponds to the minimal value that is not constrained by $\mu$ distortions \cite{Byrnes:2018txb}).\footnote{Notice that our constraints based on the mass function (\ref{eq:fPBH}) are slightly different than Ref.~\cite{Byrnes:2018txb}, which uses different definition for $f_\text{PBH}$. Their definition is based on Ref.~\cite{Niemeyer:1997mt}, which considers corrections due to large fluctuations in density. Notice that this change does not lead to any significant differences in $\sigma$.}
The solid blue line indicates when the PBH abundance coincides with 100$\%$ of the dark matter (above this the overproduction of PBHs overcloses the Universe), the red shaded region shows the bounds on $f_\text{PBH}$ from the right panel of Fig.~\ref{fig:beta}. The exponential sensitivity of PBH production on the amplitude of the scalar perturbations leads to a nearly scale-invariant bound on $A_s$ and $A_\theta$, well approximated by the overclosure bound.

In both panels we display contour lines for $\sigma^2$, characterizing the broadening of the initial GW spectrum, which is given by equations~\eqref{eq:sigma_mono} and \eqref{eq:sigma_ext}. We can see that for the monochromatic initial condition with $k_\star = [10^5, 10^{15}]~\text{Mpc}^{-1}$, we can find the maximum smearing effect in the range $\sigma^2 = [0.031, 0.060]$. For the extended configuration with $k_{\rm max}/k_{\rm min} = [10^1, 10^{9}]$, the maximum smearing can reach values in the range $\sigma^2 = [0.14, 0.66]$. For reference, the vertical green line indicates when the frequency of the scalar perturbations extends to the frequency of the peak sensitivity of LISA, $k_{\rm LISA} = 2\pi f_{\rm LISA} = 2 \cdot 10^{12}~\text{Mpc}^{-1}$. For GW spectra probed by LISA, the geometrical optics limit restricts our analysis to scalar perturbation spectra with wavenumbers well below this value. We note that within the geometric optics limit, the relevance of the maximally possible smearing of the spectrum increases logarithmically with the frequency of the GW experiment. E.g., for the LIGO experiment, the sensitivity peak is at $k_{\rm LIGO}  = 2 \cdot 10^{16}~\text{Mpc}^{-1}$.

It is important to note that for this analysis we used the linear theory. Recently, Ref.~\cite{Kalaja:2019uju} found that using non-linear calculations for the BH collapse, the PBH bounds on $\Delta_{\mathcal R}^2$ can be improved by one order of magnitude. From Eq.~\eqref{eq:smearing_numerical} , we see that this would reduce the maximally allowed value of $\sigma^2$ by an order of magnitude. Furthermore we note that larger values of $\sigma^2$ are possible if there is entropy injection from a hidden sector after the spectral deformation has occurred, diluting both GWs and PBHs. In the following, we will thus illustrate the impact of the (integrated) Sachs-Wolfe effect on some exemplary GW spectra for different values of $\sigma^2$.

\section{Observational implications}
\label{sec:Results}
\setcounter{equation}{0}

In this section we discuss observational implications of the spectral deformation of the GW spectrum.
We first discuss the deformation of a spiky spectrum (localized much more than $\sigma$ in logarithmic frequency),
and then discuss a broad spectrum.

\subsection{Spiky spectrum}
\label{subsec:Spiky}

We first consider the limiting case in which the original GW spectrum is spiky. 
This serves as a toy-model to illustrate the maximal possible effect of the smearing kernel in Eq.~\eqref{eq:Omegao_Omegas}, though concrete models with such spectra have been proposed~\cite{Saito:2009jt,Bugaev:2009zh,Bartolo:2018rku}.\footnote{In these works on GWs induced by larger scalar perturbations a sizable GW spectrum is necessarily accompanied by enhanced scalar perturbations of the same length scale. Accounting for these scalar perturbations would require an analysis beyond the geometrical optics limit, which is beyond the scope of this paper. }
This also helps us understand how the frequency reshuffling occurs for each frequency bin.
We model the spectrum with the Gaussian shape
\begin{align}
\Delta_h^{2 {\rm (s)}} (f)
&=
\frac{\Delta_{h,*}^2}{\sqrt{2 \pi \varepsilon^2}}
\exp
\left[
- \frac{(\ln \! f - \ln \! f_*)^2}{2 \varepsilon^2}
\right] \,.
\label{eq:Delta2o_spiky0}
\end{align}
This spectrum reduces to the $\delta$ function in $\varepsilon \to 0$ limit.
The resulting $\Delta_h^{2 {\rm (o)}}$ from Eq.~(\ref{eq:Delta2o_Delta2s}) becomes
\begin{align}
\Delta_h^{2 {\rm (o)}} (f)
&=
\frac{\Delta_{h,*}^2}{\sqrt{2 \pi (\varepsilon^2 + \sigma^2)}}
\left[
1 + \frac{\sigma^2}{\varepsilon^2 + \sigma^2} \, \bias \, (\ln \! f - \ln \! f_*)
\right]
\exp
\left[
- \frac{(\ln \! f - \ln \! f_*)^2}{2 (\varepsilon^2 + \sigma^2)}
\right] \,.
\label{eq:Delta2o_spiky}
\end{align}

\begin{figure}
\center
\includegraphics[width = 0.4\textwidth]{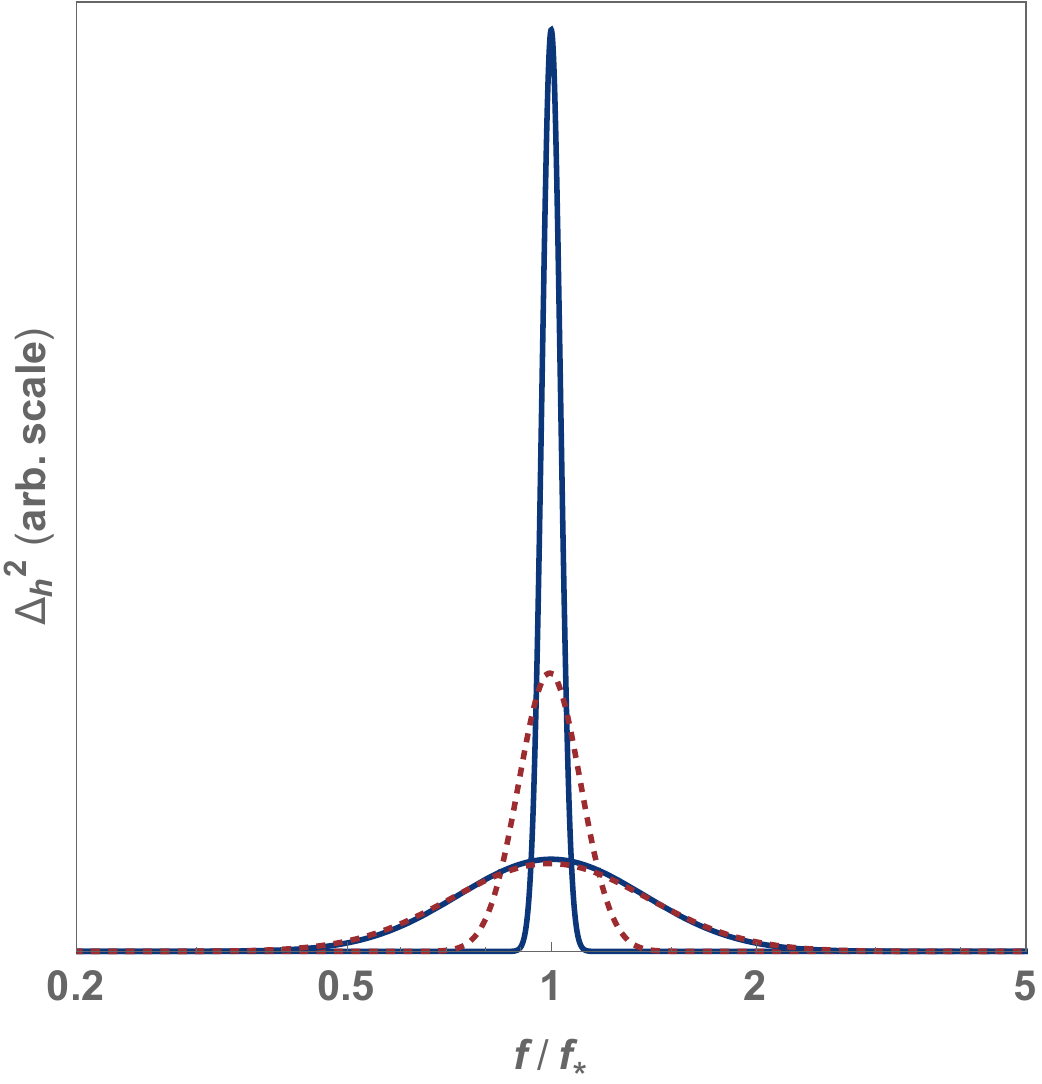}
\caption{
Comparison of the spectral deformation calculated in Eq.~(\ref{eq:Delta2o_spiky})
for $\varepsilon \ll \sigma$ and $\varepsilon \gg \sigma$.
The solid blue lines are the original spectrum (\ref{eq:Delta2o_spiky0}),
while the dotted red lines are the deformed one (\ref{eq:Delta2o_spiky}).
We fix $\sigma^2 = 0.01$, and take $\varepsilon^2 = 0.001$ and $0.1$ for the peaked and broad Gaussians, respectively.
}
 \label{fig:spiky_schematic}
\end{figure}

We plot the resulting distortion of the spectrum in Fig.~\ref{fig:spiky_schematic} for the two limiting cases $\epsilon \ll \sigma$ and $\epsilon \gg \sigma$.
To quantify the distortion, we consider the width, the maximal value and the bias of the spectrum. Comparing Eqs.~\eqref{eq:Delta2o_spiky0} and \eqref{eq:Delta2o_spiky}, the width characterizing the horizontal broadening of the spectrum changes as $\epsilon \mapsto \sqrt{\epsilon^2 + \sigma^2}$ while the decrease in the maximal amplitude, describing the leakage of the central frequency bin into the neighbouring frequency bins, is given by $1/\sqrt{2 \pi \epsilon^2} \mapsto 1/\sqrt{2 \pi (\epsilon^2 + \sigma^2)}$. The respective relative changes are thus
\begin{align}
 \text{max.\ amplitude, width:} \qquad  \sqrt{\frac{\epsilon^2}{\epsilon^2 + \sigma^2}} \simeq  \begin{cases}
                                                                        \epsilon/\sigma \quad & \text{for  } \epsilon \ll \sigma \\
                                                                       1 - \sigma^2/(2 \epsilon) \quad & \text{for  } \epsilon \gg\sigma
                                                                       \end{cases} \,,
\end{align}
illustrating that significant changes only occur for $\sigma > \epsilon$.  
To quantify the bias, we evaluate the second term in the square brackets of Eq.~\eqref{eq:Delta2o_spiky} at $|\ln \! f - \ln \! f_*| \simeq \sqrt{\sigma^2 + \epsilon^2}$, i.e.\ around the transition to the Gaussian tail. This yields
\begin{align}
 \text{bias: } \quad \frac{b \sigma^2}{\sqrt{\epsilon^2 + \sigma^2}} \simeq  b \times \begin{cases}
                                                                       \sigma \quad & \text{for  } \epsilon \ll \sigma \\
                                                                        \sigma^2/\epsilon \quad & \text{for  } \epsilon \gg\sigma
                                                                       \end{cases} \,.
\end{align}
Therefore the spectral deformation is ${\mathcal O} (\sigma)$ 
for a sufficiently localized spectrum ($\varepsilon \lesssim \sigma$)
(in the sense that the relative changes in the width, maximal amplitude and bias are all linear in $\sigma$),
while we need a more careful analysis for a broad spectrum $\varepsilon \gg \sigma$,
as we will see in Sec.~\ref{subsec:broad}.

\begin{figure}
\begin{center}
\includegraphics[width=0.4\columnwidth]{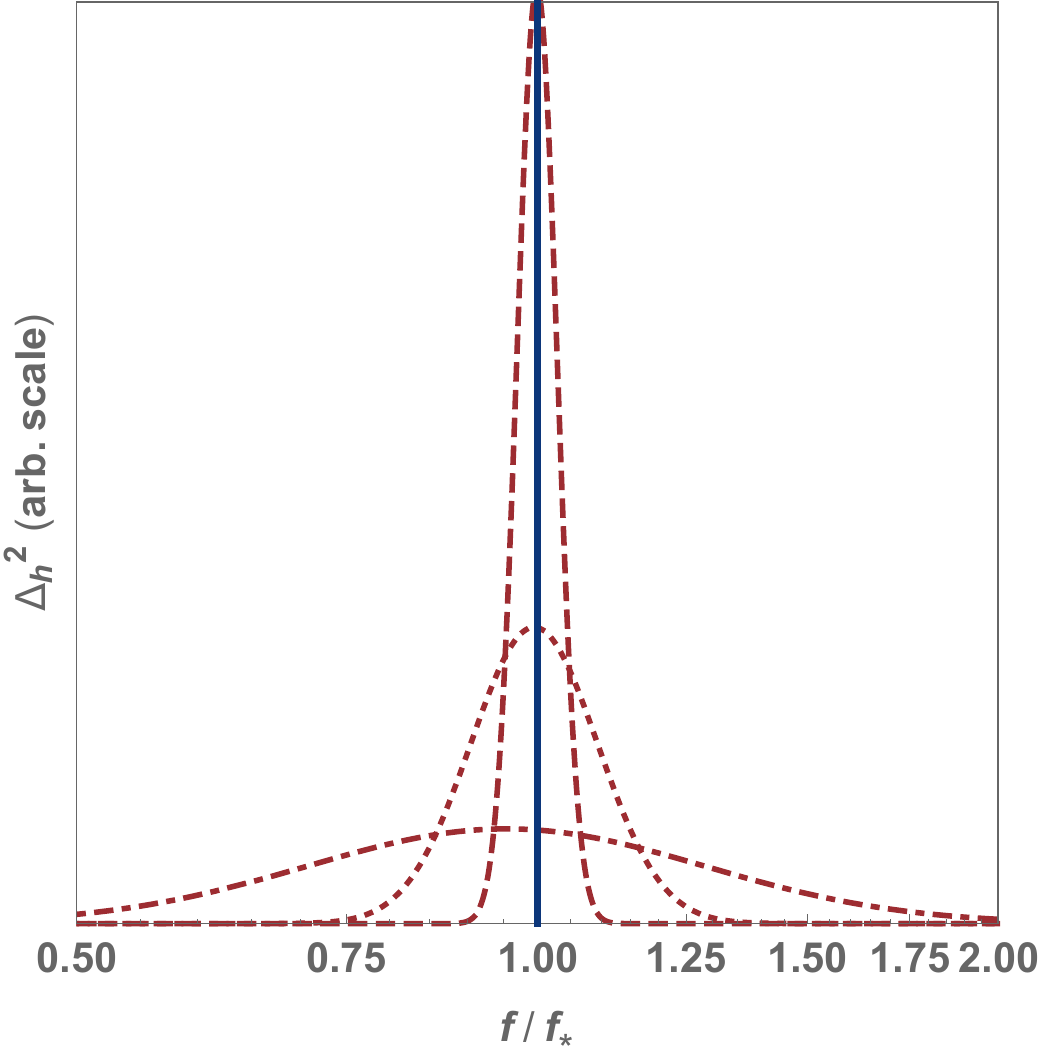} 
\hskip 0.6cm
\includegraphics[width=0.46\columnwidth]{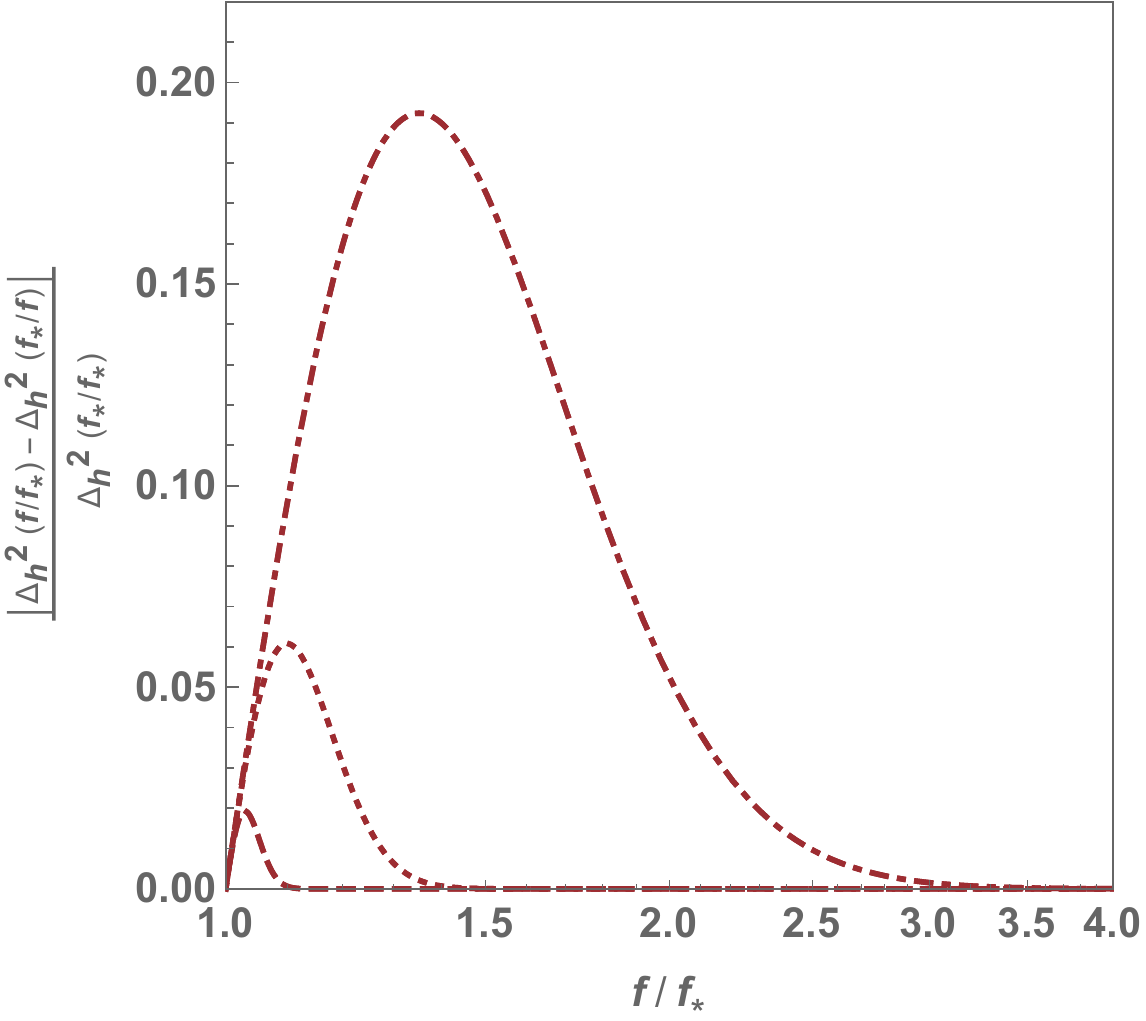}
\caption{\small 
Spectral deformation of a spiky GW spectrum.
\textit{Left panel:} 
Original (solid blue) and broadened (red) spectrum (with $\varepsilon \to 0$). 
The dashed, dotted, and dot-dashed red lines are for 
$\int d\ln \! k \, \Delta_{\mathcal R}^2 = 0.001$, $0.01$, and $0.1$, respectively.
\textit{Right panel:} 
Asymmetry of the spectrum 
$|\Delta_h^{2 {\rm (o)}} (f / f_*) - \Delta_h^{2 {\rm (o)}} (f_* / f)|$ in the left panel 
normalized by the height $\Delta_h^{2 {\rm (o)}} (f = f_*)$.
}
\label{fig:spiky}
\end{center}
\end{figure}

Fig.~\ref{fig:spiky} illustrates the distortion (left panel) and the bias towards lower frequencies (right panel) for $\epsilon \rightarrow 0$ for different values of $\int d\ln \! k \, \Delta_{\mathcal R}^2$. In agreement with the discussion above, we observe a spectral broadening and a bias of order $\sigma \sim \sqrt{\int d\ln \! k \, \Delta_{\mathcal R}^2}$.
We note that the spectral deformation is ${\mathcal O}(10)\%$ in the horizontal direction
for the maximal value of $\int d\ln \! k \, \Delta_{\mathcal R}^2$ 
allowed by the PBH constraint, $\int d\ln \! k \, \Delta_{\mathcal R}^2 \lesssim 0.4$.

We also comment on the frequency resolution of GW detectors.
The maximal frequency resolution of a GW detector is given by its observation period $T$ as $\Delta f_\text{bin} = 1/T$. An observation period of one year thus yields $\Delta f_\text{bin} \simeq 3 \cdot 10^{-8}$~Hz. Technical requirements may enforce cutting the data stream into shorter chunks, \textit{e.g.}\ a day would correspondingly yield $\Delta f_\text{bin} \simeq10^{-5}$~Hz. We conclude that for GWs with $ f \gtrsim$~mHz, the detector resolution is not a fundamental obstacle.

\subsection{Broad spectrum}
\label{subsec:broad}

In a realistic setup, we expect a broader spectrum than the one discussed in the previous subsection.
Such a broad GW spectrum can arise for example from cosmic inflation,
cosmic strings, preheating or a first-order phase transition, see Ref.~\cite{Caprini:2018mtu} for a review.

As we saw in the previous subsection, our deformation kernel (\ref{eq:kernel}) leads to 
a reshuffling of the frequency with variance $\sigma \sim \sqrt{\int d\ln \! k \, \Delta_{\mathcal R}^2}$ 
for each logarithmic frequency bin.
However, in the case of continuous GW spectra, the change in the spectrum for a fixed frequency is ${\mathcal O} (\sigma^2)$.
This can be seen explicitly by \textit{e.g.}\ considering a source spectrum approximated with the power law around $f = f_*$
\begin{align}
\Delta_h^{2 \rm (s)} (f)
&=
\Delta_{h,*}^{2 \rm (s)}
\left(
\frac{f}{f_*}
\right)^n \,,
\end{align}
for which the convolution with the smearing kernel gives
\begin{align}
\Delta_h^{2 \rm (o)} (f)
&=
\Delta_{h,*}^{2 \rm (s)}
\left(
\frac{f}{f_*}
\right)^n
\times
\left(
1 - \bias \, n \sigma^2
\right) 
e^{\frac{n^2 \sigma^2}{2}}
=
\Delta_h^{2 \rm (s)} (f)
\times
\left[
1 + {\mathcal O} (\sigma^2)
\right] \,.
\end{align}
This can be interpreted as a cancellation between the leakage from and into the fixed frequency bin.

We can also see this in more generality from a careful inspection of Eq.~(\ref{eq:Delta2o_Delta2s_general}).
Let us decompose the changes in the amplitude and frequency into different orders in the scalar perturbations
\begin{align}
\Delta \! \ln \! A
&=
\Delta \! \ln \! A^{(1)}
+
\Delta \! \ln \! A^{(2)}
+ \cdots \,,
\\
\Delta \! \ln \! f
&=
\Delta \! \ln \! f^{(1)}
+
\Delta \! \ln \! f^{(2)}
+ \cdots \,,
\end{align}
where the superscript $(i)$ denotes $i$-th order in the scalar modes.
For a broad spectrum we can expand the observed spectrum as
\begin{align}
\Delta_h^{2 {\rm (o)}} (\ln \! f)
&=
\left<
e^{2 \Delta \! \ln \! A}
\Delta_h^{2 {\rm (s)}} 
\left( \ln \! f - \Delta \! \ln \! f \right)
\right>_{\rm ens(s)}
\nonumber \\[0.1cm]
&=
\left<
\left(
1 + 2 \Delta \! \ln \! A^{(1)}
\right)
\Delta_h^{2 {\rm (s)}} 
\left( \ln \! f- \Delta \! \ln \! f^{(1)} \right)
\right>_{\rm ens(s)}
\label{eq:term1}
\\[0.1cm]
&~~~~
+
\left<
2 \left( \Delta \! \ln \! A^{(1)} \right)^2 + 2 \Delta \! \ln \! A^{(2)}
\right>_{\rm ens(s)}
\Delta_h^{2 {\rm (s)}} (\ln \! f)
\label{eq:term2}
\\[0.1cm]
&~~~~
+
\left<
\Delta_h^{2 {\rm (s)}} 
\left( \ln \! f - \Delta \! \ln \! f^{(2)} \right)
\right>_{\rm ens(s)}
-
\Delta_h^{2 {\rm (s)}} (\ln \! f)
\label{eq:term3}
\\[0.2cm]
&~~~~
+ {\mathcal O} (\sigma^3) \,.
\end{align}
We have discussed the effect of the term (\ref{eq:term1}) and seen that its effect can be written in a simple form
involving a linearly biased Gaussian kernel with a clear physical interpretation.
This term gives a dominant frequency reshuffling of order $\sigma \sim \sqrt{\int d\ln \! k \, \Delta^2_{\mathcal R}}$
(as seen from $\Delta \! \ln \! f^{(1)}$ in the argument)
compared to the second term (\ref{eq:term2}) (which just gives an overall shift in amplitude)
and the third term (\ref{eq:term3}) (which induces ${\mathcal O}(\sigma^2)$ frequency reshuffling).
This is why the term (\ref{eq:term1}) gives the dominant effect 
when the source spectrum is localized within $\sigma$ in logarithmic frequency,
as we saw in the previous subsection.
However,  since $\left< \Delta \! \ln \! A^{(1)} \right>_{\rm ens(s)} = 0 = \left< \Delta \! \ln \! f^{(1)} \right>_{\rm ens(s)}$,
the change in $\Delta_h^2$ for a fixed observed frequency 
induced by the term (\ref{eq:term1}) starts from second order in $\sigma$,
which is the same order induced by the terms (\ref{eq:term2}) and (\ref{eq:term3}).
Therefore, we need to take account of second order contributions in $\Delta \! \ln \! A$ and $\Delta \! \ln \! f$
in order to fully pin down the shape of the deformed spectrum.

\begin{figure}
\begin{center}
\includegraphics[width=0.4\columnwidth]{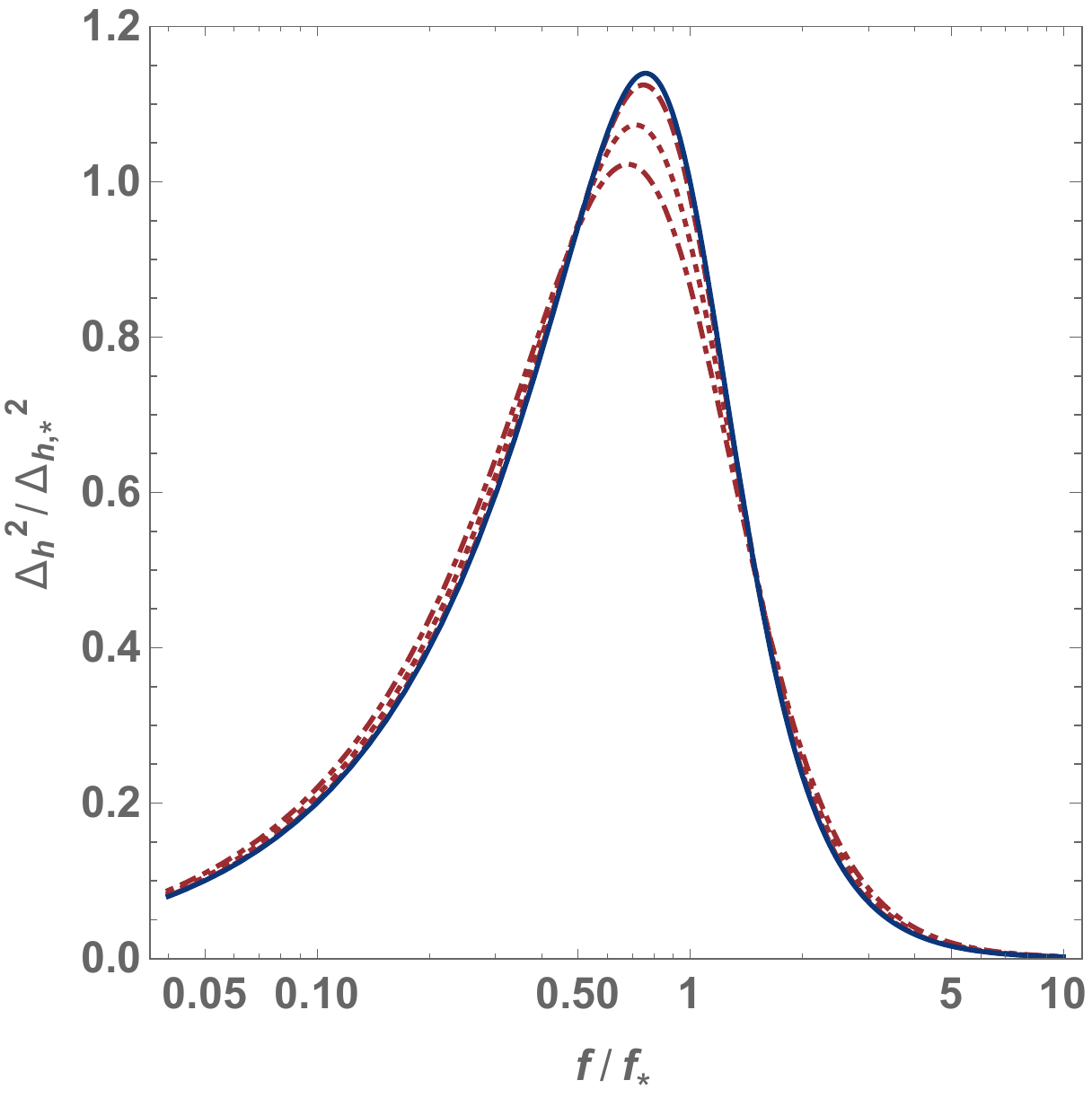}
\caption{\small
Deformation of the source spectrum (\ref{eq:broken}) with $(n_L, n_H) = (1, -3)$ calculated from the kernel (\ref{eq:kernel}). 
The dashed, dotted, and dot-dashed red lines are for 
$\int d\ln \! k \, \Delta_{\mathcal R}^2 = 0.01$, $0.05$, and $0.1$, respectively.
Note however that we need to take account of second order terms in the scalar perturbations 
appearing in amplitude and frequency changes. See the discussion in Sec.~\ref{subsec:broad}.
}
\label{fig:broken}
\end{center}
\end{figure}

With this said, it is also true that there is no {\it a priori} reason for the three terms 
(\ref{eq:term1})--(\ref{eq:term3}) to cancel out.
Therefore, to give an impression of the expected magnitude of the distortion, we show the deformation of a broad spectrum, calculated from the kernel (\ref{eq:kernel}), in Fig.~\ref{fig:broken}.
The source spectrum is chosen as
\begin{align}
\Delta_h^{2 {\rm (s)}} (f)
&=
\frac{\Delta_{h,*}^{2 {\rm (s)}}}{(f / f_*)^{- n_L} + (f / f_*)^{- n_H}}
\propto
\Delta_{h,*}^{2 {\rm (s)}}
\times
\left\{
\begin{matrix}
(f / f_*)^{n_L} 
&~~(f \ll f_*)
\\[0.1cm]
(f / f_*)^{n_H}
&~~(f \gg f_*)
\end{matrix}
\right. \,.
\label{eq:broken}
\end{align}
This spectrum has spectral indices $n_L$ and $n_H$ for low and high frequencies, respectively.
We take $(n_L, n_H) = (1, -3)$ as an example
(which arises e.g.\ from bubble collisions with thin-wall and envelope approximations~\cite{Huber:2008hg,Weir:2016tov,Jinno:2016vai,Jinno:2017fby,Konstandin:2017sat,Cutting:2018tjt} in first-order phase transitions).
However, we emphasize that all the second order contributions in $\Delta \! \ln \! A$ and $\Delta \! \ln \! f$
must be taken into account in order to accurately quantify the deformation.

\section{Discussion and conclusions}
\label{sec:DC}
\setcounter{equation}{0}

In this paper we point out the impact of density perturbations on the isotropic component of a stochastic gravitational wave background sourced at cosmological distances. With special emphasis on the Sachs-Wolfe effect and the integrated Sachs-Wolfe effect, we show that the linearized scalar perturbations at the sourcing time and along the line-of-sight modify the amplitude and frequency of the propagating GW in a correlated way. Consequently, we can estimate the observed GW spectrum by the convolution of the primordial GW spectrum with a linearly biased Gaussian kernel. The variance of this kernel function is determined by the integrated scalar power spectrum, $\sigma^2 \simeq \int d\ln \! k \, \Delta_{\mathcal R}^2$, whereas the bias encodes the correlation between frequency and amplitude changes and is fully determined by propagation equation for the GW and the background geometry.

{Equipped with this result, we can immediately quantify the impact on narrow spectra (whose width in logarithmic frequency is small compared to $\sigma$).  For a scale invariant scalar power spectrum normalized to the measured value at CMB scales, the effect is negligibly small. A strong enhancement of the scalar power spectrum on small scales, poorly constrained by current observations, can however leave a significant trace in the observed GW spectrum. Extending these result to broader spectra, we give an estimate of the expected magnitude of the deformation, but argue that a quantitative calculation requires the inclusion of higher order corrections (in terms of the scalar perturbations) in the GW propagation, which is beyond the scope of this paper. The reason for this obstacle is that to ${\mathcal O}(\sigma)$, a frequency reshuffling within an approximately flat spectrum does not alter its shape. This immediately implies that distortion effects will be suppressed by a factor of ${\mathcal O}(\sigma^2)$ for broad spectra.}

Our results are particularly relevant for future detectors while simultaneously demonstrating that such a distortion will be negligibly small for the initial discovery of the SGWB, expected in the near future. A precise determination of the spectral shape of the SGWB will, at best, be achievable with the next generation of GW detectors. The high frequency band, \textit{i.e.} Hz and beyond, is particularly relevant for early cosmological sources, entailing a long time of propagation during which scalar perturbations on many different scales enter the horizon. If not appropriately considered, sizable propagation effects could then lead to an incorrect reconstruction of the model parameters within a given cosmological model. We note that quite often, cosmological models which produce sizable tensor fluctuations also simultaneously produce sizable scalar fluctuations, which consequently will deform the GW spectrum. Typically these two types of fluctuations have a similar frequency, and their interaction can hence not be analyzed in the geometrical optics limit. We hope that our results trigger further investigation in this direction.

{The analysis presented here is subject to several simplifying assumptions. Most importantly, our analysis only includes corrections to the amplitude and frequency of propagating GWs to linear order in the scalar perturbations, and we also insist on a separation of scales between the GW frequency and the scalar perturbations. Moreover, we restricted ourselves to Gaussian scalar perturbations and neglected the Doppler contribution.} In this sense, the present work should be seen as a proof of principle of the distortion of GW spectra by density perturbations, leaving several interesting open avenues to refine in the future.

\section*{Acknowledgment}

We thank Alba Kalaja and Mauro Pieroni for helpful discussions. 
We are particularly grateful to 
Nicola Bartolo,
Daniele Bertacca,
Gabriele Franciolini,
Sabino Matarrese, 
Ilia Musco,
Marco Peloso,
Angelo Ricciardone,
Tomo Takahashi and
Gianmassimo Tasinato
for providing valuable comments on the manuscript. 
The work of RJ is supported by Grants-in-Aid for JSPS Overseas Research Fellow (No. 201960698). This work is supported by the Deutsche Forschungsgemeinschaft under Germany's Excellence Strategy -- EXC 2121 ,,Quantum Universe`` -- 390833306.

\appendix

\section{Derivation of the spectral deformation}
\label{app:Derivation}
\setcounter{equation}{0}

In this appendix we derive Eq.~\eqref{eq:Sec2_kernel}, describing the deformation of the GW spectrum.
Our assumptions are
\begin{itemize}
\item
FRW background with vanishing spatial curvature.
\item
adiabatic perturbations.
\item
separation of scales between the horizon size at the GW ``sourcing time"
(in the sense used in the main text, that is, the time when they become sufficiently sub-horizon)
and the typical wavelength of the scalar mode (see Figs.~\ref{fig:sketch2} and \ref{fig:sketch}).
\item
negligible anisotropic stress.
\end{itemize}
We also assume that the cosmic fluid consists of a single radiation component in order to simplify the calculation.
With these assumptions, 
we calculate the effect of the long-wavelength scalar modes on the propagation of the short-wavelength GWs.
We regard the latter as infinitesimal, while we treat the former to be finite and calculate its effect at the linear order.

In the following, we first derive a general relation between the GW spectrum at the sourcing and the observation time slices,
and then study the changes in GW amplitude and frequency induced by linear order terms in scalar modes.

\subsection{GW spectrum}
\label{app:GW}

We define the frequency and amplitude of GWs following Refs.~\cite{Isaacson:1967zz,Isaacson:1968zza}.
We decompose the metric $ds^2 = g_{\mu \nu} dx^\mu dx^\nu$ as
\begin{align}
g_{\mu \nu}
&= 
\bar{g}_{\mu \nu} + h_{\mu \nu} \,,
\label{eq:App_gbar_h}
\end{align}
where $\bar{g}_{\mu \nu}$ is the background metric {\it including the scalar perturbations}.
From the assumptions stated above, $\bar{g}_{\mu \nu}$ is a slowly varying function of spacetime compared to $h_{\mu \nu}$.
To extract the two tensor degrees of freedom of the GW,
we take the Lorenz gauge $\nabla_\nu \tilde{h}^{\mu \nu} = 0$ 
with $\tilde{h}_{\mu \nu} \equiv h_{\mu \nu} - \bar{g}_{\mu \nu} h/2$ and $h \equiv h^\mu_{\,\,\, \mu}$,
where the indices are raised and lowered using $\bar{g}_{\mu \nu}$.
We then take the traceless part $\tilde{h} \equiv \tilde{h}^\mu_{\,\,\, \mu} = 0$ (implying $h = 0$ 
and $\tilde h_{\mu \nu} = h_{\mu \nu}$).
We may further impose $\tilde{h}_{\mu 0} = 0$ to fix the gauge completely 
(though we do not necessarily have to).
Now, along the line of sight from the observer, we decompose the GWs $h_{\mu \nu}$ as
\begin{align}
h_{\mu \nu} (f, \hat{n})
&= 
\sum_{\lambda = +, \times}
h^{(\lambda)} (f, \hat{n}) e^{(\lambda)}_{\mu \nu} (\hat{n})
= 
\sum_{\lambda = +, \times}
A^{(\lambda)} (f, \hat{n}) e^{i \phi^{(\lambda)} (f, \hat{n})} e^{(\lambda)}_{\mu \nu} (\hat{n}) \,.
\label{eq:App_hmunu_decompose}
\end{align}
with the polarization tensor $e^{(\lambda)}_{\mu \nu}$ normalized as 
$e^{(\lambda)}_{\mu \nu} e^{(\lambda') \mu \nu} = \delta_{\lambda \lambda'}$.
Here the unit vector $\hat{n}$ specifies the line-of-sight direction, and $f$ denotes the frequency of the GWs.
In this expression the label for time is implicit:
we consider the sourcing time (s) or observation time (o) in the following.
The real functions $A^{(\lambda)}$ and $\phi^{(\lambda)}$ give the amplitude and phase for the polarization $\lambda$.
In the following we omit the label $(\lambda)$ for simplicity.

We next derive the relation between the GW spectrum at the sourcing and observation time slices.
Let us define the changes in the logarithmic amplitude and frequency $\Delta \! \ln \! A$ and $\Delta \! \ln \! f$ as
(see App.~\ref{app:step1} for a precise definition of the sourcing time slice)
\begin{align} 
\Delta \! \ln \! A
&\equiv 
\ln (a_{\rm o} A_{\rm o} / a_{\rm s} A_{\rm s}) \,,
\\
\Delta \! \ln \! f
&\equiv 
\ln (a_{\rm o} f_{\rm o} / a_{\rm s} f_{\rm s}) \,.
\end{align}
In the following we omit the scale factor whenever the redshift in FRW background is trivial.
From Eq.~(\ref{eq:h_decomposition}), GWs at the two time slices are related as
\begin{align} 
h_{\rm o} 
\left( 
f_{\rm o}, \hat{n}_{\rm o} 
\right)
f_{\rm o}^2 df_{\rm o}
&=
e^{\Delta \! \ln \! A}~
h_{\rm s}
\left(
f_{\rm s}, \hat{n}_{\rm s}
\right)
f_{\rm s}^2 df_{\rm s} \,,
\end{align}
where the frequencies $f_{\rm o}$ and $f_{\rm s}$ are related as $f_{\rm o} = e^{\Delta \! \ln \! f} f_{\rm s}$.
The propagation directions $\hat{n}_{\rm o}$ and $\hat{n}_{\rm s}$ are also related (via lensing)
as $\hat{n}_{\rm o} = \hat{n}_{\rm s} + \Delta \hat{n}$.
As a result, we have 
\begin{align}
&
\left<
h_{\rm o} (f, \hat{n})
h_{\rm o}^* (f', \hat{n}')
\right>_{\rm ens(s,t)} 
\nonumber \\
&=
\left<
\frac{}{}
e^{\Delta \! \ln \! A} ~
e^{\Delta \! \ln \! A'}~
e^{-3\Delta \! \ln \! f}~
e^{-3\Delta \! \ln \! f'}
\right.
\nonumber \\
&~~~~~~
\left.
\times
\left<
h_{\rm s} 
\left(
e^{-\Delta \! \ln \! f} f, \hat{n} - \Delta \hat{n}
\right)
h_{\rm s}^*
\left(
e^{-\Delta \! \ln \! f'} f', \hat{n}' - \Delta \hat{n}'
\right)
\right>_{\rm ens(t)}
\right>_{\rm ens(s)}
\nonumber \\
&=
\left<
\frac{}{}
e^{2 \Delta \! \ln \! A}~
e^{-6 \Delta \! \ln \! f}
\right.
\nonumber \\
&~~~~~~
\left.
\times
\left<
h_{\rm s} 
\left(
e^{-\Delta \! \ln \! f} f, \hat{n} - \Delta \hat{n}
\right)
h_{\rm s}^*
\left(
e^{-\Delta \! \ln \! f} f', \hat{n}' - \Delta \hat{n}'
\right)
\right>_{\rm ens(t)}
\right>_{\rm ens(s)} \,.
\label{eq:App_hh}
\end{align}
Here $\langle \cdots \rangle_{\rm ens(s)}$ and $\langle \cdots \rangle_{\rm ens(t)}$ denote
the scalar and tensor ensemble averages, respectively.
We assume that both averages are independent.
In the first equality $\Delta \! \ln \! A = \Delta \! \ln \! A (\hat{n}, \left\{ {\mathcal R}_{\rm pr} \right\})$
and $\Delta \! \ln \! A' = \Delta \! \ln \! A (\hat{n}', \left\{ {\mathcal R}_{\rm pr} \right\})$ are understood.
In the second equality we replaced $\hat{n}'$ with $\hat{n}$ except where $\delta^2 (\hat{n}, \hat{n}')$ appears later.
From Eqs.~(\ref{eq:hh_Ph}) and (\ref{eq:Deltah2_Ph}) 
we can relate the ensemble average to the GW power spectrum as
\begin{align}
\left<
h_{\rm o} (f, \hat{n})
h_{\rm o}^* (f', \hat{n}')
\right>_{\rm ens(s,t)} 
&\propto
\frac{1}{f^5}~
\delta (f - f')~
\delta^2 (\hat{n}, \hat{n}')~
\Delta_h^{2 {\rm (o)}}(f) \,,
\\
\left<
h_{\rm s} (f'', \hat{n} - \Delta \hat{n})
h_{\rm s}^* (f''', \hat{n}' - \Delta \hat{n}')
\right>_{\rm ens(t)} 
&\propto
\frac{1}{f''^5}~
\delta (f'' - f''')~
\delta^2 (\hat{n}, \hat{n}')~
\Delta_h^{2 {\rm (s)}}(f'') \,.
\end{align}
Therefore, taking into account the scaling dimensions of the delta function in frequency space, 
the overall factor of $e^{-6 \Delta \! \ln \! f}$ in Eq.~(\ref{eq:App_hh}) cancels out 
when we write the deformation in terms of the power spectrum:
\begin{align}
\Delta_h^{2 {\rm (o)}}(f)
&=
\left<
e^{2 \Delta \! \ln \! A}
\Delta_h^{2 {\rm (s)}} 
\left( e^{-\Delta \! \ln \! f} f \right)
\right>_{\rm ens(s)}.
\label{eq:App_Delta2o_Delta2s_general}
\end{align}

Now, since the changes in the amplitude and frequency come from the scalar perturbations at the intermediate scales,
we can expand $\Delta \! \ln \! A$ and $\Delta \! \ln \! f$ as
\begin{align} 
\Delta \! \ln \! A
&=
\Delta \! \ln \! A^{(1)} + \Delta \! \ln \! A^{(2)} + \cdots \,,
\\
\Delta \! \ln \! f
&=
\Delta \! \ln \! f^{(1)} + \Delta \! \ln \! f^{(2)} + \cdots \,,
\end{align}
with the superscript $(i)$ denoting $i$-th order in the scalar perturbations.
These quantities denote changes along the line of sight of the propagating GW,
but we can safely replace them with the scalar ensemble average, as we see below.
In this paper we study the effect of first order terms $\Delta \! \ln \! A^{(1)}$ and $\Delta \! \ln \! f^{(1)}$
on the deformation of the GW spectrum.
With a slight abuse of notation, $\Delta_h^2 (\ln \! f) = \Delta_h^2 (f)$, to simplify expressions below, 
we study the implications of the equation
\begin{align}
\Delta_h^{2 {\rm (o)}} (\ln \! f)
&\simeq
\left<
\left( 1 + 2 \Delta \! \ln \! A^{(1)} \right)
\Delta_h^{2 {\rm (s)}} 
\left( \ln \! f - \Delta \! \ln \! f^{(1)} \right)
\right>_{\rm ens(s)} \,.
\label{eq:App_Delta2o_Delta2s_linear}
\end{align}
For a monochromatic GW spectrum or for each individual frequency bin of a broader spectrum, this expression accurately describes the deformation.
See Sec.~\ref{sec:Results} for the discussion on the effect of the $\Delta \!\ln \!A^{(2)}$ and $\Delta \!\ln \!f^{(2)}$ terms.

\begin{figure}
\centering{
\includegraphics[width = 0.8\textwidth]{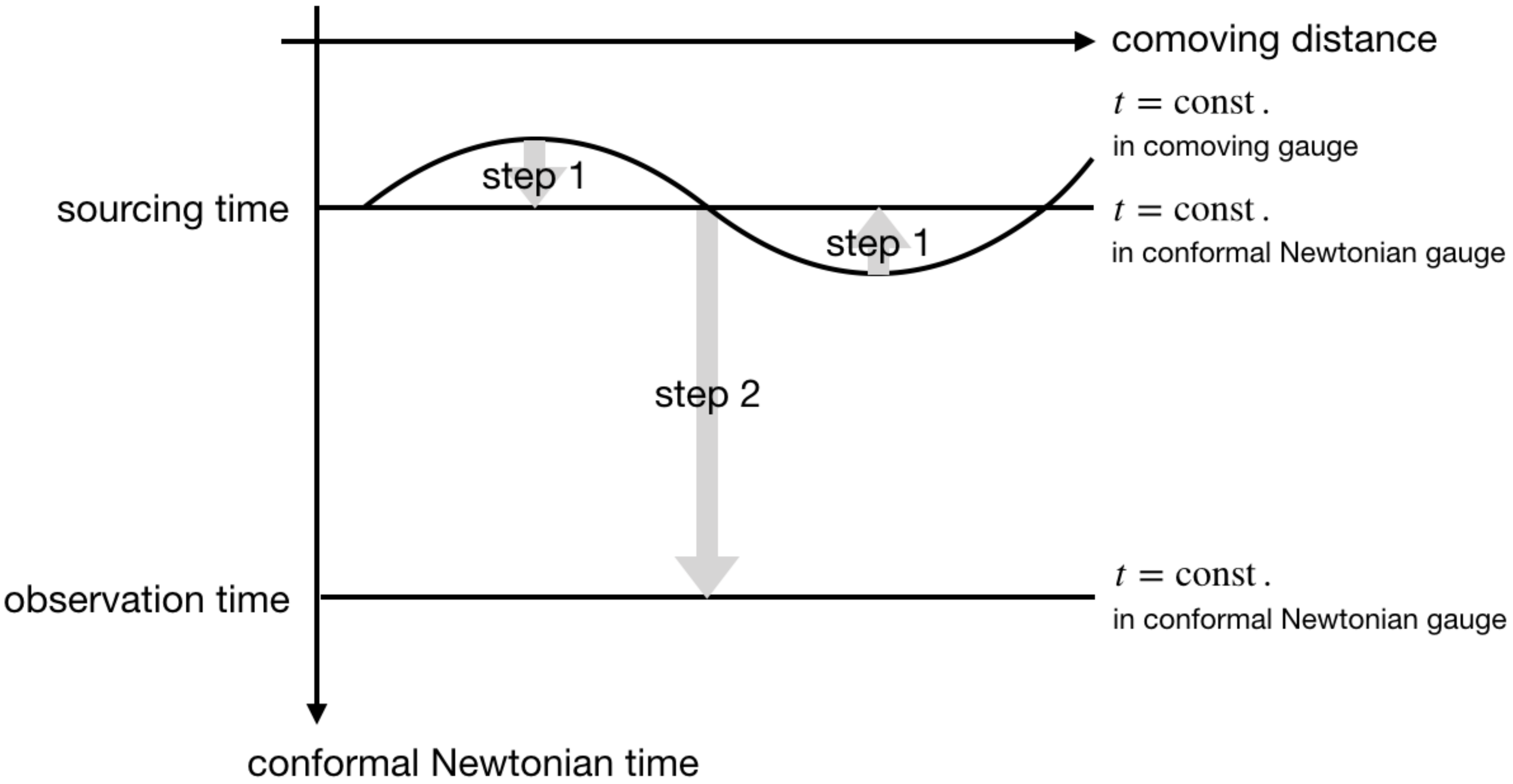} 
}
\caption{
The two steps in the derivation of the deformation equation.
}
\label{fig:gauge}
\end{figure}

We derive Eq.~(\ref{eq:kernel}) for the Gaussian kernel from 
Eq.~(\ref{eq:App_Delta2o_Delta2s_linear}).
Our derivation proceeds in two steps.
We first identify the GW spectra typically computed in the literature 
(calculated assuming a flat, homogeneous FRW Universe)
as the one on the constant-time hypersurface in so-called 
comoving~\cite{Hu:1998tj} or velocity-orthogonal isotropic~\cite{Kodama:1985bj} gauge (hereafter called comoving gauge),
and calculate the changes in the GW frequency and amplitude
when we move from this gauge to the conformal Newtonian gauge (step 1 in Fig.~\ref{fig:gauge}, subscript ``init'' below).
We start from the former gauge because the density fluctuation vanishes and the fluid proper time coincides 
with the coordinate time in the large scale limit.
This procedure basically follows the derivation of the SW effect in Ref.~\cite{White:1997vi}.
We then calculate the propagation effect from the initial to the final hypersurface 
in the conformal Newtonian gauge (step 2 in Fig.~\ref{fig:gauge}, subscript ``prop'' below).
We use the results of Ref.~\cite{Laguna:2009re} on the frequency and amplitude change for  GWs from point sources.

\subsection{Step 1: GW spectrum at the sourcing time}
\label{app:step1}

This subsection closely follows Ref.~\cite{White:1997vi},
in which the SW effect is derived by moving from the comoving gauge to the conformal Newtonian gauge.
The former is defined by the conditions ${\mathcal B} = v$ and ${\mathcal H}_T = 0$ in the general decomposition of the metric 
\begin{align}
\bar{g}_{\mu \nu} dx^\mu dx^\nu
&=
a^2 \left[
- (1 + 2 {\mathcal A}) d\tau^2 - {\mathcal B}_i d\tau dx^i 
+ \left[ (1 + 2 {\mathcal H}_L) \gamma_{ij} + 2 {\mathcal H}_{T ij} \right] dx^i dx^j
\right] \,,
\end{align}
and the energy-momentum tensor
\begin{align}
T^0_{\,\,\,\, 0}
&=
- (\rho + \delta \rho) \,,
~~
T^0_{\,\,\,\, i}
=
-(\rho + p) (v_i - {\mathcal B}_i) \,,
~~
T_0^{\,\,\,\, i}
=
(\rho + p) v_i \,,
~~
T^i_{\,\,\, j}
=
(p + \delta p) \delta^i_{\,\,\,j} + p \Pi^i_{\,\,\, j} \,.
\end{align}
Here $d\tau \equiv adt$ is the conformal time, 
and all the three-quantities are raised and lowered with the spatial metric $\gamma_{ij}$, 
which is equal to $\delta_{ij}$ in our setup.
Also, the scalar components are defined like
${\mathcal A} \supset {\mathcal A} (\vec{k}) Q^{(0)}(\vec{k})$, ${\mathcal B}_i \supset {\mathcal B} (\vec{k}) Q_i^{(0)}(\vec{k})$,
${\mathcal H}_{T ij} \supset {\mathcal H}_T(\vec{k}) Q_{ij}^{(0)}(\vec{k})$ and so on using the dimensionless eigenfunctions
defined through $(\gamma^{ij} \nabla_i \nabla_j  - k^2) Q^{(0)} = 0$, $Q_i^{(0)} = - (\nabla_i / k) Q^{(0)}$,
and $Q_{ij}^{(0)} = (\nabla_i \nabla_j / k^2 + \gamma_{ij} / 3) Q^{(0)}$ with $k^2 \equiv k_i k^i$.

Useful properties of this gauge are
(1) the density fluctuation $\delta \rho$ (and thus $\delta p$) vanishes in the $k \to 0$ limit,
and
(2) the proper time of the fluid coincides with the coordinate time in the $k \to 0$ limit.
Indeed, the Poisson equation
\begin{align}
k^2 {\mathcal H}_L + k H {\mathcal B}
&= 
4 \pi G \delta \rho \,,
\end{align}
gives $\delta \rho \to 0$ for $k \to 0$ with $H$ being the Hubble parameter.
Also, the Euler equation 
\begin{align}
(\rho + p) {\mathcal A}
&= 
- \delta p + \frac{2}{3} p \Pi \,,
\end{align}
gives ${\mathcal A} \to 0$ for $k \to 0$ in the absence of the anisotropic stress $\Pi$.
Therefore, we identify the standard GW spectrum, derived in the FRW background, 
as the one at the constant-time hypersurface in this comoving gauge.

Now let us see how the GW frequency and amplitude change 
when we move to the conformal Newtonian gauge.
Since in our setup the Newtonian potential is effectively constant inside each Hubble patch at the sourcing time,
the relation between the time variables in the conformal Newtonian (CN) and the comoving (com) gauges is given by
\begin{align}
dt_{\rm com}
&= 
(1 + \Phi_{\rm s}) dt_{\rm CN}
~~
\to
~~
t_{\rm com}
= 
(1 + \Phi_{\rm s}) t_{\rm CN} \,.
\end{align}
Note that this relation applies at a fixed spacetime point.
This means that temperature fluctuations exist on the equal-time hypersurface in the conformal Newtonian gauge.
Using $a \propto t^{1/2}$ in radiation domination, and also using the fact that GWs redshift as a radiation component,
we have
\begin{align}
(\Delta \! \ln \! A^{(1)})_{\rm init}
&\equiv 
\left[
\ln \! A_{\rm s, CN} - \ln \! A_{\rm s, com}
\right]^{(1)}
=
- \frac{1}{2} \Phi_{\rm s} \,,
\label{eq:App_DeltalnA1}
\\
(\Delta \! \ln \! f^{(1)})_{\rm init}
&\equiv 
\left[
\ln \! f_{\rm s, CN} - \ln \! f_{\rm s, com}
\right]^{(1)}
= 
- \frac{1}{2} \Phi_{\rm s} \,,
\label{eq:App_Deltalnf1}
\end{align}
on the initial conformal Newtonian hypersurface.\footnote{
Note that ``s,com" and ``s,CN" refer to different points along the time direction.
}

\subsection{Step 2: Evolution from the sourcing time to the observation time}
\label{app:step2}

In this subsection, we discuss how perturbations affect the GW propagation through the (I)SW effect and through lensing. 
This subsection closely follows Refs.~\cite{Isaacson:1967zz,Laguna:2009re}.

Adopting the geometric optics limit, 
we decompose $h$ into the amplitude $A$ and the phase $\phi$ as in Eq.~(\ref{eq:App_hmunu_decompose}).
The GW wave vector $k_\mu$ is a vector normal to surfaces of constant phase 
\begin{align}
k_\mu = \nabla_\mu\phi \,.
\end{align}
Then the equation of motion for the GWs gives $k_\mu k^\mu = 0$.
We introduce an integration curve parametrized by $l$ as
\begin{align}
\frac{dx^\mu}{dl}
&= k^\mu \,,
\end{align}
then $l$ becomes an affine parameter because 
$\nabla_\nu (k_\mu k^\mu) = 2 (\nabla_\nu k_\mu) k^\mu = 2 k^\mu (\nabla_\mu k_\nu) = 0$ implies $d k_\nu / dl = 0$.
Next we use the fact that null geodesics with the background metric $\bar{g}_{\mu \nu}$ with the affine parameter $l$
are the same as the ones with the rescaled metric $\tilde{\bar{g}}_{\mu \nu} \equiv \bar{g}_{\mu \nu} / a^2$ 
with the affine parameter $d\lambda = dl / a$ and the wave vector $\tilde{k}^\mu = a^2 k^\mu$.
Writing the derivative with respect to $\lambda$ as $d_\lambda \equiv d/d\lambda = a k^\mu\nabla_\mu$,
we have
\begin{align} 
\frac{1}{a}~d_\lambda \ln \! A^{(1)}
&= - \frac{1}{2} \nabla_a k^a \,.
\label{eq:App_amplitude_change}
\end{align}
We expand the rescaled wave vector $\tilde{k}^\mu$ in terms of the scalar perturbations
\begin{align}
 \tilde{k}^\mu(\lambda) = \tilde{k}^{(0)\mu}(\lambda) + \tilde{k}^{(1)\mu}(\lambda) \,,
\end{align}
and take $\tilde{k}^{(0) 0} = 1$ unit in the following, 
since we are interested only in the fractional shift in the frequency.
The zeroth-order solution is just a constant propagation in the line-of-sight direction 
$\hat{n}$ with $\tilde{k}^{(0) \mu} = (1,-n^i)$, 
and the first-order solution is given by the geodesic equation
\begin{align} 
d_\lambda \tilde{k}^{(1)\mu} + \tilde{\Gamma}^{(1)\mu}_{\nu \rho}\tilde{k}^{(0)\nu}\tilde{k}^{(0)\rho} 
&= 0 \,.
\label{eq:App_geodesic}
\end{align}
Adopting the conformal Newtonian gauge as in Eq.~\eqref{eq:metric}, 
we have the following expressions for the connection
\begin{align}
\tilde{\Gamma}^{(1)0}_{00} = \Phi' \,,
~~~~
\tilde{\Gamma}^{(1)0}_{0i} = \partial_i \Phi \,,
&~~~~
\tilde{\Gamma}^{(1)i}_{00} = \partial^i \Phi \,,
~~~~
\\
\tilde{\Gamma}^{(1)0}_{ij} = -\Psi' \delta_{ij} \,,
~~~~
\tilde{\Gamma}^{(1)i}_{0j} = -\Psi' \delta^i_{\,\,\, j} \,,
&~~~~
\tilde{\Gamma}^{(1)i}_{jk} = \left( \delta_{jk} \partial^i - \delta^i_{\,\,\, j} \partial_{k} - \delta^i_{\,\,\, k} \partial_{j} \right) \Psi \,,
\end{align}
where the prime denotes the conformal time derivative.
Denoting $\tilde{k}^{(1)i}_\parallel = n^i n_j \tilde{k}^{(1) j}$ and 
$\tilde{k}^{(1) i}_\bot = ( \delta^i_{\,\,\, j} - n^i n_j ) \tilde{k}^{(1) j} \equiv \bot^i_{\,\,\, j} \tilde{k}^{(1) j}$, 
and also noticing that $d_\lambda \Phi = \Phi' + n^i \partial_i \Phi$ at the unperturbed level 
(this is enough since other terms are already first order in perturbations), 
we find for the temporal and spatial parts of Eq.~\eqref{eq:App_geodesic}
\begin{align}
d_\lambda \tilde{k}^{(1)0} &= - 2d_\lambda \Phi + (\Phi' + \Psi') \,, 
\\
d_\lambda \tilde{k}^{(1)i}_\parallel &=
\left[  d_\lambda (-\Phi + \Psi) +  (\Phi' + \Psi') \right] n^i \,, 
\\
d_\lambda \tilde{k}^{(1)i}_\bot &= - \bot^i_{\,\, j} \partial^j (\Phi + \Psi) \,.
\end{align}
Integrating along the line-of-sight, we obtain
\begin{align}
\tilde{k}^{(1)0} (\lambda)
&= 
- 2\Phi|_{\lambda_{\rm s}}^\lambda - (\Phi + \Psi)|_{\lambda_{\rm s}}
+ \int_{\lambda_{\rm s}}^\lambda d\lambda' \, (\Phi' + \Psi') \,,
\\
\tilde{k}^{(1)i}_\parallel (\lambda)
&=
\left[ (-\Phi + \Psi)|_{\lambda_{\rm s}}^\lambda 
+  \int_{\lambda_{\rm s}}^\lambda d\lambda' \, (\Phi' + \Psi') \right] n^i \,,
\\
\tilde{k}^{(1)i}_\bot (\lambda)
&= 
- \bot^i_{\,\,\, j} \int_{\lambda_{\rm s}}^\lambda d\lambda' \, \partial^j (\Phi + \Psi) \,.
\end{align}

\paragraph{Amplitude change.}

From Eq.~(\ref{eq:App_amplitude_change}), we have
\begin{align}
&d_\lambda (\Delta \ln \! \tilde{A}^{(1)})_{\rm prop}
\nonumber \\
&=  
- \frac{1}{2}
\left[
\partial_\tau \tilde{k}^{(1)0}+ \partial_i \tilde{k}^{(1)i}_\parallel +\partial_i \tilde{k}^{(1)i}_\bot + \Gamma ^{(1)\mu}_{\mu\nu} \tilde{k}^{(0)\nu}
\right]
\nonumber \\
&= 
\frac{1}{2} \partial_\tau \left( \Psi+\Phi \right) 
- \frac{1}{2} d_\lambda \left[ -2\Psi+\int_{\lambda_{\rm s}}^{\lambda_o} d\lambda'\,( \Phi' + \Psi') \right] 
+ \frac{1}{2} \bot^{ij} \int_{\lambda_{\rm s}}^{\lambda_{\rm o}}  d\lambda'\,\partial_i \partial_j (\Phi + \Psi) \,.
\end{align}
The tilde on $(\Delta \! \ln \! A^{(1)})_{\rm prop}$ marks this as an auxiliary quantity, as we will be able to drop some of these contributions in our final expression, see below.
Integrating along the line-of-sight, we find 
\begin{align} 
(\Delta \ln \! \tilde{A}^{(1)})_{\rm prop}
&\equiv
\left[
\ln (a_{\rm o} A_{\rm o}) - \ln (a_{\rm s} A_{\rm s, CN})
\right]^{(1)}
\nonumber \\
&= 
(\Delta \! \ln \! A^{(1)})_{\rm SW} + (\Delta \! \ln \! A^{(1)})_{\rm lens} 
\nonumber \\
&= 
(\Psi_{\rm o} - \Psi_{\rm s}) 
+ \frac{1}{2} \bot^{ij} 
\int_{\lambda_{\rm s}}^{\lambda_{\rm o}} d\lambda
\int_{\lambda_{\rm s}}^\lambda d\lambda' \,\partial_i \partial_j (\Phi + \Psi) \,.
\end{align}
The first term can be identified as the Sachs-Wolfe contribution\footnote{
One may wonder why the amplitude decreases $(\Delta \! \ln \! A)_{\rm SW} = \Psi_{\rm o} - \Psi_{\rm s} < 0$
when the GW escapes from a positive Newtonian potential
$\Phi_{\rm s} = \Psi_{\rm s} > 0$ to the observer's position $\Phi_{\rm o} = \Psi_{\rm o} = 0$.
Here, we have to be careful about the normalization condition of the polarization tensor 
$e_{\mu \nu} e^{\mu\nu} = e_{ij} e^{ij} = 1$, which gives $e_{ij} \propto 1 - 2\Psi$ in our setup
(here we assumed the gauge condition $h_{\mu 0} = 0$).
The combination $A e_{ij}$ increases from the source to the observer:
$(\Delta \! \ln \! A \, e_{ij})_{\rm SW} = \Psi_{\rm s} - \Psi_{\rm o}$.
} 
and the second as the gravitational lensing term.
Note that the ISW term has canceled out.

Next we argue that only $(\Delta \! \ln \! A^{(1)})_{\rm SW}$ should be taken into account
when discussing the amplitude change in the power spectrum of stochastic GWs.
The lensing contribution $(\Delta \! \ln \! A^{(1)})_{\rm lens}$ simply comes from the change in the cross section of the light rays 
at the sourcing and observation points.
Indeed, parallel light rays with an infinitesimal area $S_{\rm s}$ converges to the area $S_{\rm o}$ given by
\begin{align}
S_{\rm o} 
&= 
S_{\rm s} \times \left[
1 +  \int_{\lambda_{\rm s}}^{\lambda_{\rm o}} d\lambda~
\partial_i \tilde{k}^{(1)i}_\bot (\lambda)
\right]
=
S_{\rm s} \times \left[
1 -  
\bot^{ij} 
\int_{\lambda_{\rm s}}^{\lambda_{\rm o}} d\lambda
\int_{\lambda_{\rm s}}^\lambda d\lambda' \,\partial_i \partial_j (\Phi + \Psi)
\right] \,,
\end{align}
at the leading order in the scalar perturbations.
The lensing term $(\Delta \! \ln \! A^{(1)})_{\rm lens}$ can be understood as keeping (area)$\times$(amplitude)${}^2$ constant.
This effect is just a rearranging of the propagation direction, and can be neglected when discussing the isotropic component of the stochastic GWs.
Therefore the amplitude change we need is
\begin{align}
(\Delta \! \ln \! A^{(1)})_{\rm prop}
&= 
(\Delta \! \ln \! A^{(1)})_{\rm SW}
= 
\Psi_{\rm o} - \Psi_{\rm s} \,.
\label{eq:App_DeltalnA2}
\end{align}

\paragraph{Frequency shift.}

We can calculate the frequency by contracting with the fluid velocity $u^\mu = \left( 1 - \Phi,v^i \right) / a$ as
\begin{align}
\omega (\lambda)
&= 
2 \pi f (\lambda)
= 
-u^\mu k_\mu 
= 
\frac{1}{a}\left[
1 + \hat{n} \cdot \vec{v} 
- \Phi|_{\lambda_{\rm s}}^\lambda 
- \Psi|_{\lambda_{\rm s}} 
+ \int_{\lambda_{\rm s}}^\lambda d\lambda'~(\Phi' + \Psi') 
\right] \,.
\end{align}
Therefore, the logarithmic shift in the comoving frequency $af$ is calculated as
\begin{align}
(\Delta \! \ln \! f^{(1)})_{\rm prop}
&\equiv
\left[
\ln (a_{\rm o} f_{\rm o}) - \ln (a_{\rm s} f_{\rm s, CN})
\right]^{(1)}
\nonumber \\[0.2cm]
&=
(\Delta \! \ln \! f^{(1)})_{\rm Doppler} + (\Delta \! \ln \! f^{(1)})_{\rm SW} + (\Delta \! \ln \! f^{(1)})_{\rm ISW}
\nonumber \\
&=
\hat{n} \cdot (\vec{v}_{\rm o} - \vec{v}_{\rm s}) 
- (\Phi_{\rm o} - \Phi_{\rm s}) 
+ \int_{\lambda_{\rm s}}^{\lambda_{\rm o}} d\lambda~
\partial_\tau (\Phi + \Psi) \,.
\label{eq:App_Deltalnf2}
\end{align}

\subsection{Combining the two steps}
\label{app:combining}

Now we combine the two steps Eqs.~(\ref{eq:App_DeltalnA1}) and (\ref{eq:App_Deltalnf1}), 
and Eqs.~(\ref{eq:App_DeltalnA2}) and (\ref{eq:App_Deltalnf2}),
to derive the Gaussian kernel for the deformation of the power spectrum.
From App.~\ref{app:step1} and \ref{app:step2}, we have
\begin{align} 
&\Delta \! \ln \! A^{(1)}
= 
\left[
\ln (a_{\rm o} A_{\rm o}) - \ln (a_{\rm s} A_{\rm s,com})
\right]^{(1)}
= (\Delta \! \ln \! A^{(1)})_{\rm init} + (\Delta \! \ln \! A^{(1)})_{\rm prop} \,,
\label{eq:App_DeltalnA}
\\[0.2cm]
&\Delta \! \ln \! f^{(1)}
= 
\left[
\ln (a_{\rm o} f_{\rm o}) - \ln (a_{\rm s} f_{\rm s,com})
\right]^{(1)}
= (\Delta \! \ln \! f^{(1)})_{\rm init} + (\Delta \! \ln \! f^{(1)})_{\rm prop} \,.
\label{eq:App_Deltalnf}
\end{align}
The result for the SW term in $\Delta \! \ln \! f^{(1)}$ is in particular consistent with the results derived in Refs.~\cite{Alba:2015cms,Bartolo:2019zvb}.

In the following we neglect the Doppler term and the scalar perturbations at the observer
$\Phi_{\rm o}$ and $\Psi_{\rm o}$, as explained in Sec.~\ref{sec:Assumptions}.
We rewrite Eq.~(\ref{eq:App_Delta2o_Delta2s_linear}) using a $\delta$-function as
\begin{align}
\Delta_h^{2 {\rm (o)}}(f)
&\simeq
\int d\ln \! f'~
\Delta_h^{2 {\rm (s)}}  (f')
\nonumber \\
&~~~~~~
\times
\left<
\left(
1 + 2 \Delta \! \ln \! A^{(1)} \left( \hat{n}, \left\{ {\mathcal R}_{\rm pr} \right\} \right)
\right)
\frac{}{}
\delta
\left(
\ln \! f - \ln \! f' - \Delta \! \ln \! f^{(1)} \left( \hat{n}, \left\{ {\mathcal R}_{\rm pr} \right\} \right)
\right)
\right>_{\rm ens(s)} \,.
\label{eq:App_Delta2o_Delta2s_delta}
\end{align}
The second line in Eq.~(\ref{eq:App_Delta2o_Delta2s_delta}) corresponds to the case with
a $\delta$-function source peaked around $\ln \! f'$: $\Delta_h^{2 {\rm (s)}} (\ln \! f) = \delta (\ln \! f - \ln \! f')$.
We first calculate the deformation for this $\delta$-function spectrum and then substitute the result into 
Eq.~(\ref{eq:App_Delta2o_Delta2s_delta}).

To eliminate the correlations between the two variables $\Delta \! \ln \! A^{(1)}$ and $\Delta \! \ln \! f^{(1)}$, we perform a change of basis using an orthogonal matrix:
\begin{align}
\left(
\begin{matrix}
\Delta \! \ln \! A^{(1)} \\[1ex]
\Delta \! \ln \! f^{(1)}
\end{matrix}
\right)
&=
\left(
\begin{matrix}
c & - s
\\[1ex]
s & c
\end{matrix}
\right)
\left(
\begin{matrix}
\Delta_1 \\[1ex]
\Delta_2
\end{matrix}
\right) \,,
\end{align}
with $c \equiv \cos \theta$ and $s \equiv \sin \theta$, 
and $0 \leq \theta < \pi$ is chosen so that $\Delta_1$ and $\Delta_2$ satisfy
\begin{align}
\left<
\Delta_1 \Delta_2
\right>_{\rm ens(s)}
&=
0 \,.
\label{eq:App_Delta1Delta2}
\end{align}
This condition gives the value of $\theta$ in terms of the variances of $\Delta \! \ln \! A^{(1)}$ and $\Delta \! \ln \! f^{(1)}$:
\begin{align}
\tan 2 \theta
&=
\frac{2\left< \Delta \! \ln \! A^{(1)}~\Delta \! \ln \! f^{(1)} \right>_{\rm ens(s)}}
{\left< (\Delta \! \ln \! A^{(1)})^2 \right>_{\rm ens(s)} - \left< (\Delta \! \ln \! f^{(1)})^2 \right>_{\rm ens(s)}} \,.
\end{align}
We assume the curvature perturbation $\left\{ {\mathcal R}_{\rm pr} \right\}$ to be Gaussian.
In this case, the scalar ensemble average reduces to the Gaussian integrations
with respect to $\Delta_1$ and $\Delta_2$.
Defining the variances $\sigma_1^2 \equiv \left< \Delta_1^2 \right>_{\rm ens(s)}$ 
and $\sigma_2^2 \equiv \left< \Delta_2^2 \right>_{\rm ens(s)}$, 
the spectral deformation for the $\delta$-function case is calculated as
\begin{align}
&\left<
\left( 1 + 2 \Delta \! \ln \! A^{(1)} \right)
\frac{}{}
\delta
\left(
\ln \! f - \ln \! f' - \Delta \! \ln \! f^{(1)}
\right)
\right>_{\rm ens(s)}
\nonumber \\[0.1cm]
&=
\int d \Delta_1
\int d \Delta_2~
\frac{e^{-\Delta_1^2 / 2 \sigma_1^2}}{\sqrt{2 \pi \sigma_1^2}}
~
\frac{e^{-\Delta_2^2 / 2 \sigma_2^2}}{\sqrt{2 \pi \sigma_2^2}}
~
\left(1 + 2 \Delta \! \ln \! A^{(1)} \right)
\frac{}{}
\delta
\left(
\ln \! f - \ln \! f' - \Delta \! \ln \! f^{(1)}
\right)
\nonumber \\[0.1cm]
&=
\frac{1}{\sqrt{2 \pi (s^2 \sigma_1^2 + c^2 \sigma_2^2)}}
\left[
1 + \frac{2c s (\sigma_1^2 - \sigma_2^2)}{s^2 \sigma_1^2 + c^2 \sigma_2^2}
(\ln \! f - \ln \! f')
\right]
e^{-\frac{1}{2} \frac{(\ln \! f - \ln \! f')^2}{s^2 \sigma_1^2 + c^2 \sigma_2^2}}
\nonumber \\[0.2cm]
&\equiv
\frac{1}{\sqrt{2 \pi \sigma^2}}
\left[
1 + b (\ln \! f - \ln \! f')
\right]
e^{- \frac{(\ln \! f - \ln \! f')^2}{2 \sigma^2}} \,,
\label{eq:App_Deformation_delta}
\end{align}
where
\begin{align}
&\sigma^2
\equiv
s^2 \sigma_1^2 + c^2 \sigma_2^2~
\left( = \left< (\Delta \! \ln \! f^{(1)})^2 \right>_{\rm ens(s)} \right) \,,
\\
&b
\equiv
\frac{2c s (\sigma_1^2 - \sigma_2^2)}{s^2 \sigma_1^2 + c^2 \sigma_2^2}
\left( = 2 \frac{\left< \Delta \! \ln \! A^{(1)}~\Delta \! \ln \! f^{(1)} \right>_{\rm ens(s)}}{\left< (\Delta \! \ln \! f^{(1)})^2 \right>_{\rm ens(s)}} \right) \,.
\end{align}
The final expression in Eq.~(\ref{eq:App_Deformation_delta}) is instructive:
The original $\delta$-function localized at $\ln \! f'$ is smeared by the last Gaussian factor,
while the correlation between $\Delta \! \ln \! A^{(1)}$ and $\Delta \! \ln \! f^{(1)}$ introduces a linear bias for the Gaussian.
As a result, we obtain the following deformation equation
\begin{align}
\Delta_h^{2 {\rm (o)}}(\ln \! f)
&\simeq
\int d\ln \! f'~
\Delta_h^{2 {\rm (s)}} 
\left(
\ln \! f'
\right)
\times
\frac{1}{\sqrt{2 \pi \sigma^2}}
\left[
1 + b (\ln \! f - \ln \! f')
\right]
e^{- \frac{(\ln \! f - \ln \! f')^2}{2 \sigma^2}} \,.
\label{eq:App_Delta2o_Delta2s_final}
\end{align}

Finally we calculate the numerical values of $\theta$, $\sigma_1^2$ and $\sigma_2^2$.
We write $\Delta \! \ln \! A^{(1)}$ and $\Delta \! \ln \! f^{(1)}$ as
\begin{align}
\Delta \! \ln \! A^{(1)}
&=
\Delta \! \ln \! A^{(1)} (\hat{n}, \left\{ {\mathcal R}_{\rm pr} \right\})
=
\int_k
{\mathcal R}_{\rm pr} (\vec{k}) {\mathcal A}(\vec{k}, \hat{n}) \,,
\label{eq:App_DeltalnA_decomposed}
\\
\Delta \! \ln \! f^{(1)}
&=
\Delta \! \ln \! f^{(1)} (\hat{n}, \left\{ {\mathcal R}_{\rm pr} \right\})
=\int_k
{\mathcal R}_{\rm pr} (\vec{k}) {\mathcal F}(\vec{k}, \hat{n}) \,,
\label{eq:App_Deltalnf_decomposed}
\end{align}
where $\int_k \equiv \int d^3k / (2 \pi)^3$.
From Eqs.~(\ref{eq:App_DeltalnA})--(\ref{eq:App_Deltalnf}), and from the transfer function~\cite{BaumannCosmology}
\begin{align}
\Phi (\tau, \vec{k})
&=
\Psi (\tau, \vec{k})
= 
\frac{2}{3} T (k\tau) {\mathcal R}_{\rm pr} (\vec{k}) \,,
~~~~
T (k\tau)
=
\frac{9}{k^2 \tau^2}
\left[
\frac{\sin (k \tau / \sqrt{3})}{k \tau / \sqrt{3}} - \cos(k \tau / \sqrt{3})
\right] \,,
\end{align}
we find that ${\mathcal A}$ and ${\mathcal F}$ have the following forms
\begin{align}
{\mathcal A}(\vec{k}, \hat{n})
&=
-e^{- i (\vec{k} \cdot \hat{n}) (\tau_{\rm o} - \tau_{\rm s})}~
T(k \tau_{\rm s}) \,,
\\[0.2cm]
{\mathcal F}(\vec{k}, \hat{n})
&=
\frac{1}{3}~
e^{- i (\vec{k} \cdot \hat{n}) (\tau_{\rm o} - \tau_{\rm s})}~
T(k \tau_{\rm s})
+
\frac{2}{3}
\int_{\tau_{\rm s}}^{\tau_{\rm o}} d\tau~
e^{- i (\vec{k} \cdot \hat{n}) (\tau_{\rm o} - \tau)}
\partial_\tau T(k \tau) \,,
\end{align}
where we used $\Phi = \Psi$.
Since $k \tau_{\rm s} \ll 1 \ll k \tau_{\rm o}$ holds in our setup,
${\mathcal A}$ and ${\mathcal F}$ are simplified as
\begin{align}
{\mathcal A}(\vec{k}, \hat{n})
&\simeq
-e^{- i c_k (k \tau_{\rm o} - k \tau_{\rm s})} \,,
\\[0.2cm]
{\mathcal F}(\vec{k}, \hat{n})
&\simeq
e^{- i c_k (k \tau_{\rm o} - k \tau_{\rm s})}
\left[
\frac{1}{3} 
+ 
\frac{2}{3} \int_0^\infty d (k \tau)~
e^{i c_k (k \tau)}~
T'(k \tau)
\right] \,,
\end{align}
with $c_k \equiv \hat{k} \cdot \hat{n} \equiv (\vec{k} / k) \cdot \hat{n}$. 
Note that the quantities in the square brackets depend only on $c_k$.
Using these expressions, we numerically obtain
\begin{align}
\left<
(\Delta \! \ln \! A^{(1)})^2
\right>_{\rm ens(s)}
&\simeq 
1.00
\times
\int d\ln \! k~
\Delta_{\mathcal R}^2 \,,
\\
\left<
\Delta \! \ln \! A^{(1)}~\Delta \! \ln \! f^{(1)}
\right>_{\rm ens(s)}
&\simeq 
- 0.240
\times
\int d\ln \! k~
\Delta_{\mathcal R}^2 \,,
\\
\left<
(\Delta \! \ln \! f^{(1)})^2
\right>_{\rm ens(s)}
&\simeq 
0.914
\times
\int d\ln \! k~
\Delta_{\mathcal R}^2 \,.
\end{align}
As a result, the rotation angle $\theta$ is calculated as $\theta \simeq -0.696$,
and the variance $\sigma$ and the linear bias $b$ in Eq.~(\ref{eq:App_Delta2o_Delta2s_final}) become
\begin{align}
\sigma^2
&\simeq
0.914
\times
\int d\ln \! k~
\Delta_{\mathcal R}^2 \,,
~~~~~~
b
\simeq
-0.524 \,.
\end{align}

\small
\bibliography{ref}

\end{document}